%% file: main.tex
\documentclass[fleqn,10pt]{wlscirep}
\input{preamble.tex}

\title{Adaptive Segmentation of Knee Radiographs for Selecting the Optimal ROI in Texture Analysis}
\author[1]{Neslihan Bayramoglu}
\author[1,3]{Aleksei Tiulpin}
\author[2]{Jukka Hirvasniemi}
\author[1,3,4]{Miika T. Nieminen}
\author[1,3,4]{Simo Saarakkala}
\affil[1]{Research Unit of Medical Imaging, Physics and Technology (MIPT), University of Oulu, Finland}
\affil[2]{Department of Radiology and Nuclear Medicine, Erasmus MC, University Medical Center, Rotterdam, The Netherlands}
\affil[3]{Department of Diagnostic Radiology, Oulu University Hospital, Oulu, Finland
}
\affil[4]{Medical Research Center, University of Oulu and Oulu University Hospital, Oulu, Finland}

\begin{abstract}
    \input{sections/abstract.tex}

\end{abstract}

\begin{document}
\maketitle
\thispagestyle{empty}

\input{sections/introduction.tex}
\input{sections/materials_and_methods.tex}

\input{sections/results_conclusion.tex}

\printbibliography

\newpage
\input{sections/supplement.tex}

\end{document}

%% file: preamble.tex
\usepackage[utf8]{inputenc}
\usepackage{setspace}

\usepackage{graphicx}
\usepackage{subfiles}
\usepackage{authblk}
\usepackage{caption}
\usepackage{rotating}

\usepackage{booktabs,array}
\usepackage{colortbl}
\newcolumntype{P}[1]{>{\centering\arraybackslash}p{#1}}

\usepackage{amssymb}
\usepackage{amsmath}
\usepackage{hyperref}
\usepackage{subfig}

\usepackage{geometry}

\newcommand{\textBf}{\fontseries{b}\selectfont}

\hypersetup{
    colorlinks=true,
    citecolor=blue,
    linkcolor=blue,
    filecolor=magenta,      
    urlcolor=blue}

\usepackage{nameref}

\usepackage{multirow}

\usepackage[
backend=biber,
bibstyle=nature,
citestyle=numeric-comp,
sorting=none,
natbib=true,
url= true, 
doi=true,
eprint=true,
autocite=superscript,
backref=true,
]{biblatex}

\DefineBibliographyStrings{english}{%
    backrefpage  = {on p.}, % for single page number
    backrefpages = {on pp.} % for multiple page numbers
}

\usepackage{filecontents}

\let\cite=\supercite
\graphicspath{{images/}{../images/}}

%% file: sections/abstract.tex
\paragraph{Objective:}
The purposes of this study were to investigate: 1) the effect of placement of region-of-interest (ROI) for texture analysis of subchondral bone in knee radiographs, and 2) the ability of several texture descriptors to distinguish between the knees with and without radiographic osteoarthritis (OA).

\paragraph{Design:}
Bilateral posterior-anterior knee radiographs were analyzed from the baseline of Osteoarthritis Initiative (OAI) and Multicenter Osteoarthritis Study (MOST) datasets.
A fully automatic method to locate the most informative region from subchondral bone using adaptive segmentation was developed.
We used an oversegmentation strategy for partitioning knee images into the compact regions that follow natural texture boundaries.
Local Binary Patterns (LBP), Fractal Dimension (FD), Haralick features, Shannon entropy, and Histogram of Oriented Gradients (HOG) methods were computed within the standard ROI and within the proposed adaptive ROIs. Subsequently, we built logistic regression models to identify and compare the performances of each texture descriptor and each ROI placement method using 5-fold cross validation setting. Importantly, we also investigated the generalizability of our approach by training the models on OAI and testing them on MOST dataset. We used area under the receiver operating characteristic (ROC) curve (AUC) and average precision (AP) obtained from the precision-recall (PR) curve to compare the results.

\paragraph{Results:} 
We found that the adaptive ROI improves the classification performance (OA vs. non-OA) over the commonly-used standard ROI (up to 9\% percent increase in AUC). We also observed that, from all texture parameters, LBP yielded the best performance in all settings with the best AUC of 0.840 [0.825, 0.852] and associated AP of 0.804 [0.786, 0.820].

\paragraph{Conclusion:} Compared to the current state-of-the-art approaches, our results suggest that the proposed adaptive ROI approach in texture analysis of subchondral bone can increase the diagnostic performance for detecting the presence of radiographic OA.


%% file: sections/introduction.tex
\section{Introduction}

Along with the progression of osteoarthritis (OA), subchondral bone (SB) is subject to changes in its structure and composition \cite{buckland2004subchondral,kamibayashi1995trabecular}.
In knee radiography, a strong association between SB texture and severity of OA has been reported 
\cite{lynch1991analysis, janvier2017subchondral,hirvasniemi2017differences,buckland2004subchondral,ljuhar2018combining,jennane2014variational,janvier2017subchondral,kraus2018predictive,hirvasniemi2014quantification,wolski2010differences,thomson2015automated,podsiadlo2016baseline,messent2005tibial,kraus2009trabecular}. In particular, the thickness of subchondral cortical plate is increased especially in the medial compartment knee OA \cite{buckland2004subchondral}. Moreover, ladder-like appearance is observed in horizontal trabeculae in OA \cite{buckland2004subchondral}.
Although the thickness and volume of subchondral bone increases in OA, it is weaker and less mineralized than normal bone\cite{buckland2004subchondral}.
It is also known that trabecular network adapts to alterations in joint loading relatively quickly \cite{lowitz2014characterization}.

Plain radiography (X-ray imaging) is a cheap and widely available clinical modality to detect the presence and severity of OA \cite{tiulpin2018automatic}.  
Unfortunately, plain radiography has generally relatively low sensitivity for detecting early osteoarthritic changes \cite{neogi2012clinical}. Further, the status of SB is rarely assessed from plain radiographs in clinical practice, and instead, the overall severity of OA within a joint is visually evaluated or semi-quantitatively graded. However, it has been proposed that the quantification of SB structural changes with texture analysis from radiographs can lead to development of more sensitive OA biomarkers \cite{hirvasniemi2014quantification, lowitz2014characterization}.

Various studies have focused on quantitative analysis of knee joint radiographs \cite{tiulpin2019multimodal, tiulpin2018automatic, brahim2019decision, minciullo2016fully, antony2016quantifying, thomson2015automated}, and in particular on SB texture analysis \cite{hirvasniemi2019bone, kraus2018predictive, hirvasniemi2014quantification,riad2018texture, woloszynski2012prediction, podsiadlo2014trabecular}.
Changes in SB texture in the radiograph have been quantified based on roughness, anisotropy, and orientation of texture elements mostly by fractal methods \cite{lynch1991analysis, janvier2017subchondral,hirvasniemi2017differences,buckland2004subchondral,ljuhar2018combining,jennane2014variational,kraus2018predictive,hirvasniemi2014quantification,wolski2010differences,thomson2015automated,podsiadlo2016baseline,messent2005tibial,kraus2009trabecular, wolski2009directional,janvier2015roi, thomson2015automated}.
Only a few studies have investigated non-fractal methods for classifying SB texture \cite{shamir2009knee, thomson2015automated,hladuuvka2017femoral,hirvasniemi2014quantification,riad2018texture,anifah2013osteoarthritis}. 
The potential advantage of using non-fractal methods is to capture statistical characteristics of textures and added discriminative power by distinguish key texture primitives such as edges, corners and uniform regions \cite{varma2007locally}.
The placement of region-of-interest (ROI) is a crucial step in bone texture analysis. Most often it is done  manually \cite{lynch1991robust, messent2006differences,kraus2009trabecular, hirvasniemi2014quantification, woloszynski2010signature}, but there exist some studies where semi-automatic\cite{janvier2017subchondral,hladuuvka2017femoral} and fully automatic approaches \cite{thomson2015automated, podsiadlo2008automated} have been used for the ROI placement. It is well known that bony changes in OA are typically not uniform throughout the subchondral bone, but instead highly localized regional differences in bone microstructure can be observed \cite{kamibayashi1995trabecular}. Consequently, placement of ROI could have a significant effect on the image analysis results, and thus, selecting ROI for texture analysis plays an important role to accurately quantify changes in subchondral bone structure.

\begin{figure}[t]
\centering
\subfloat[Medial side ROI]{\includegraphics[height = 6cm]{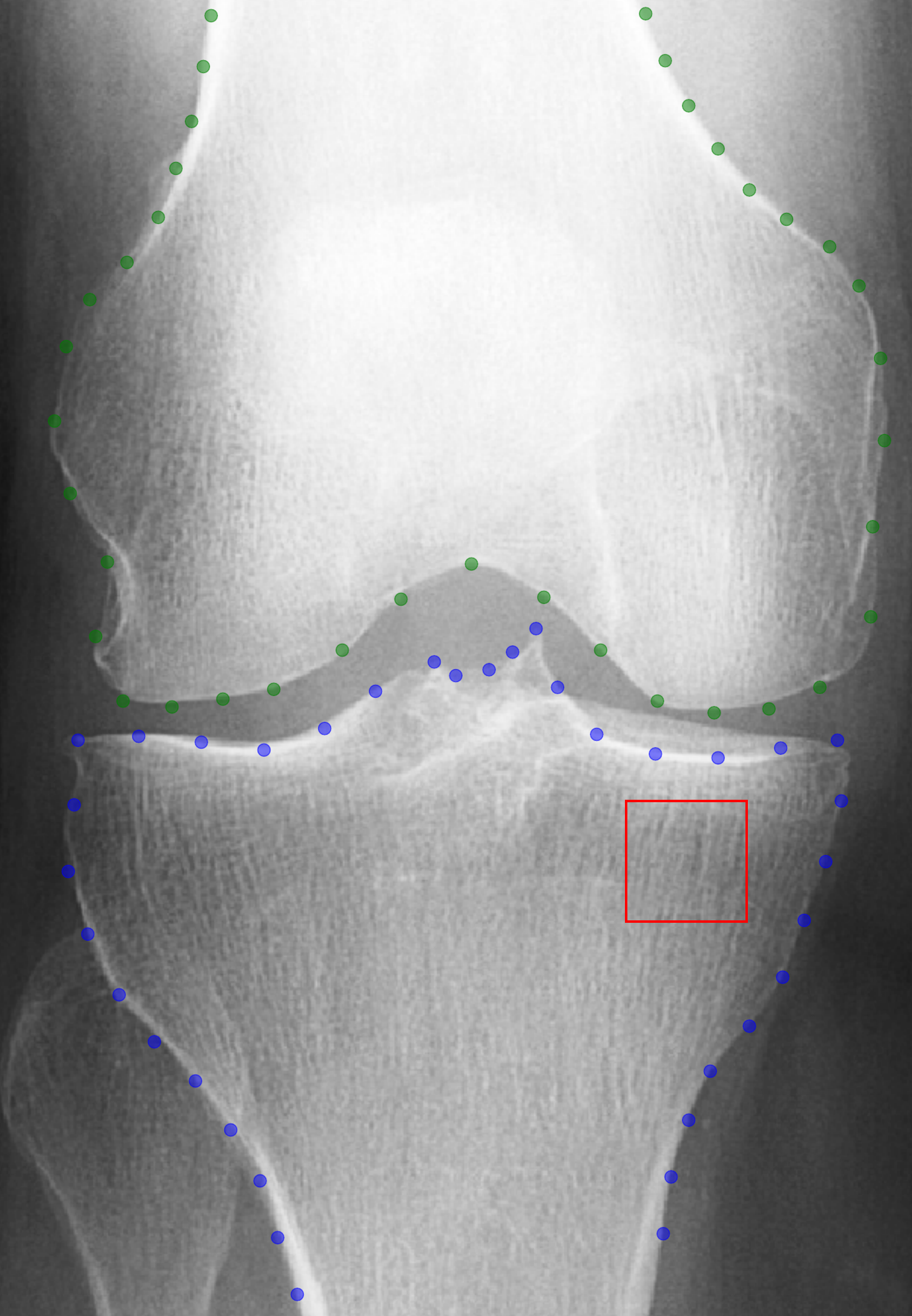}}%
 \hfill
\subfloat[Medial and Lateral ROIs]{\includegraphics[height = 6cm]{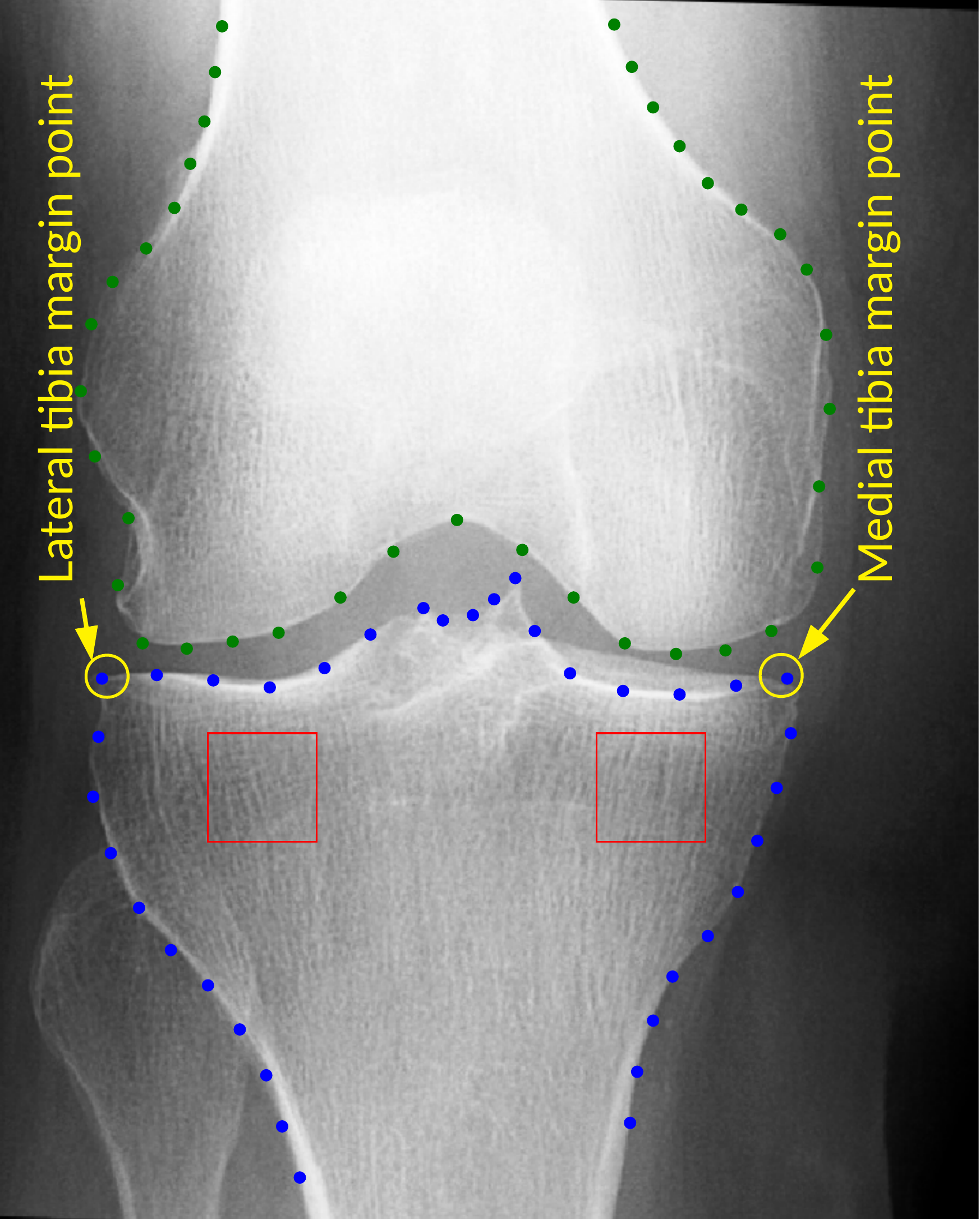}}%
\hfill
\subfloat[ROIs arranged in a lattice]{\includegraphics[height = 6cm]{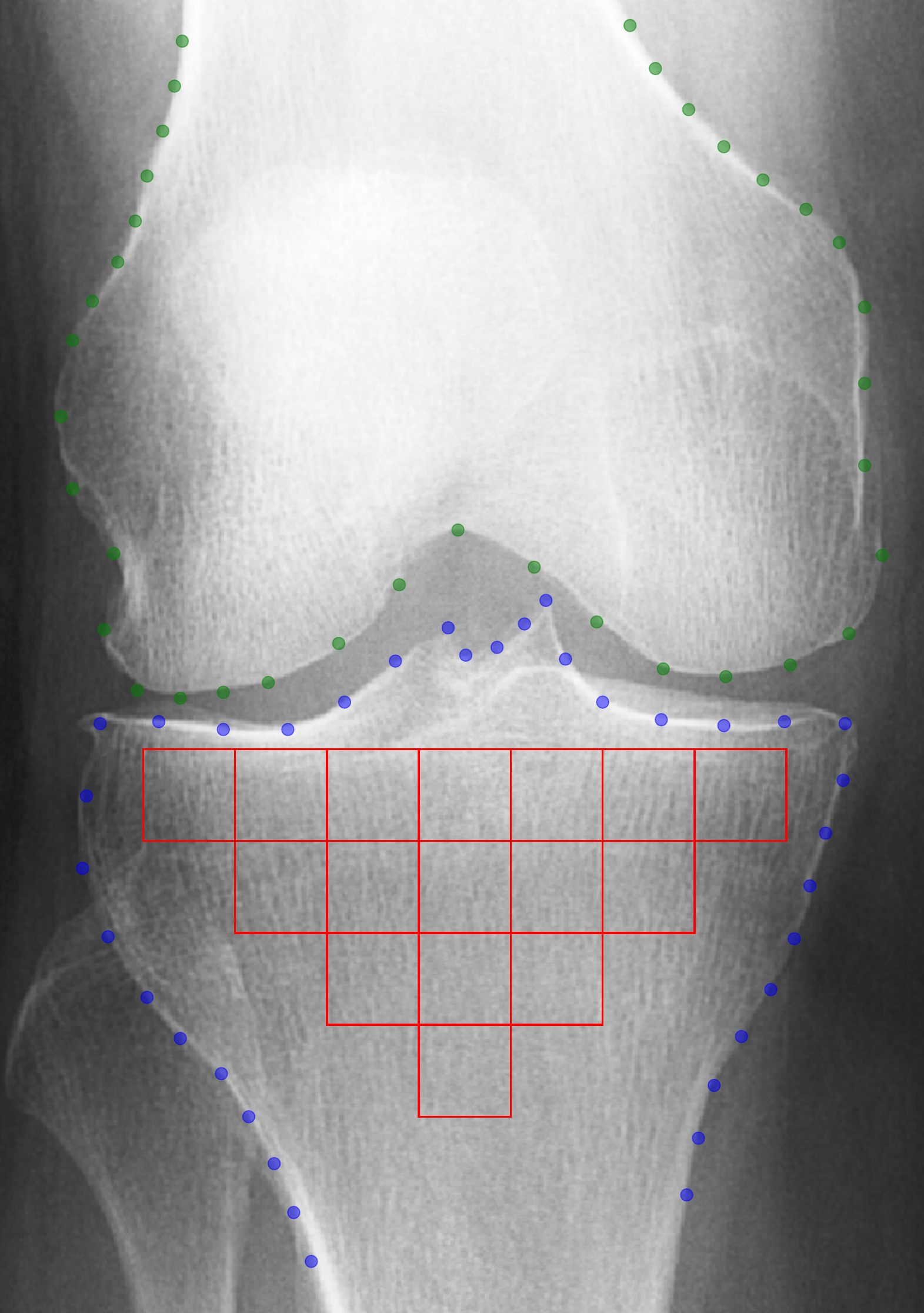}}%
\caption{Examples of region-of-interest (ROI) placement for texture analysis in previous studies.
Here, a) shows an example of medial side ROI, b) shows medial and lateral ROIs together with medial and lateral tibia margins, and c) square ROIs arranged in a lattice. Typical landmark points are also marked in the images.}
\label{fig: roi}
\end{figure}

Earlier studies have used predefined rectangular ROIs either at a single location or at multiple sites \cite{messent2006differences, ljuhar2018combining, podsiadlo2008automated,janvier2015roi,hirvasniemi2014quantification,hladuuvka2017femoral,kraus2009trabecular, buckland2004subchondral} (see Figure \ref{fig: roi}).
In particular, rectangular ROIs are often placed immediately beneath the inferior border of the medial or lateral cortical plates \cite{lynch1991analysis,kraus2009trabecular,hafezi2018new,messent2005tibial,hirvasniemi2017differences,podsiadlo2008automated,hirvasniemi2014quantification}.
Most of the approaches found in the literature employ medial ROI as medial OA has a much higher prevalence in the population.
Previously, the outer regions of the tibial compartment have been excluded from the analysis intentionally to avoid the inclusion of periarticular osteopenia adjacent to marginal osteophytes\cite{messent2005tibial}.
However, only a few studies have questioned the effects of the placement, size, and shape of ROI on the prediction of knee OA in texture based methods \cite{janvier2015roi,hladuuvka2017femoral}. Specifically, Janvier et al. \cite{janvier2015roi} demonstrated the importance of the ROI placement by arranging square ROIs in a lattice using fractal descriptors (See Figure \ref{fig: roi}). In their study, the square side length was equal to $1/7$ of the tibial width minus an offset of $10\%$ to prevent the periarticular malformations. They concluded that the success of predicting incident OA depends on the location of the ROI.

\begin{figure}[htp]
\centering
\includegraphics[width = 1\linewidth, trim=1mm 0mm 0mmin 0mm,clip]{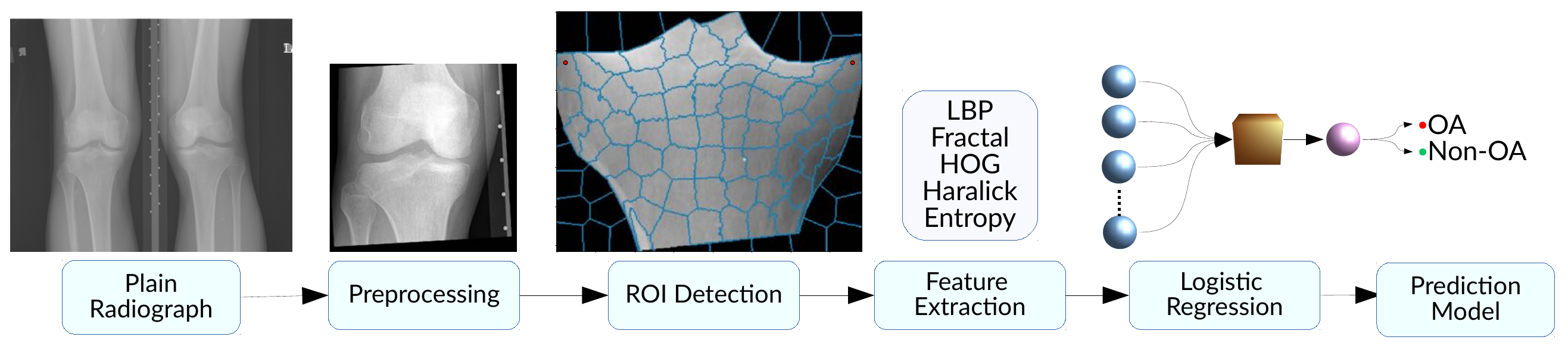}
\caption{Flowchart of our classification pipeline for the detection of radiographic OA. As a first step, we localized bone landmarks using BoneFinder software (see Materials and Methods for more details). Subsequently, we applied global contrast normalisation and histogram truncation between the $5^{th}$ and $99^{th}$ percentiles.
We then normalised the data to have a 0.2mm pixel spacing. Finally, each knee is rotated in order to have an aligned horizontal tibial plateau. After ROI detection and texture description, we applied logistic regression in a 5-fold cross validation setting to measure the performance of all the methods.}
\label{fig: flowchart}
\end{figure}

In this study, we propose a fully automated method to locate the most informative subchondral bone ROI in plain radiographs using adaptive segmentation. We conducted extensive analyses on the placement of texture ROI for both femur and tibia, including their outer regions, and comparing the results using several texture descriptors using cross-validation and an independent test set. 
The flowchart of the proposed pipeline is illustrated in Figure \ref{fig: flowchart}. The method is based on oversegmentation and it partitions the knee images into compact regions that respect the local texture boundaries.
We believe that such division is more natural than employing fixed rectangular ROIs because texture boundaries do not follow straight lines, but are instead following true anatomical structure of the bone.

%% file: sections/materials_and_methods.tex
\section{Materials and Methods}

\subsection{Data}
\label{sec:data}

We used the data  from the Osteoarthritis Initiative (OAI) and Multicenter Osteoarthritis Study (MOST).
OAI comprises individuals at risk of development of symptomatic femorotibial OA.
A total of 4,796 men and women at ages 45–79 were enrolled to the study between 2004 and 2006.
Images were graded by Kellgren–Lawrence (KL) scoring system.
We selected all the knees at baseline that were graded by KL score on both knees thereby excluding the cases with total knee replacement.
Multicenter Osteoarthritis Study (MOST) comprises 3,026 individuals aged 50–79 years who either had radiographic knee OA or were at high risk for developing the disease.
At the baseline clinical visit, all the subjects underwent weight-bearing posteroanterior fixed flexion knee radiographs.
From MOST data, we selected all the subjects at baseline with posteroanterior view having acquired with 10 degrees beam angle (PA10) of the knee given that both joints are graded with KL score. 
Knees with total knee replacement were excluded.
The details about OAI and MOST datasets used in this study are presented in Table \ref{tab: data}.
Future details about OAI and MOST can be found on \href{http://www.oai.ucsf.edu/ }{http://www.oai.ucsf.edu/} and  \href{http://most.ucsf.edu}{http://most.ucsf.edu}, respectively.

In our analysis, we trained algorithms on OAI dataset and tested them on MOST data which are independent study materials.
Such a validation is very important to asses the methods' performance objectively and its ability to analyze unseen data.

\begin{table}[htb]
\caption{Description of the data. 
We used the data  from the OAI dataset and MOST dataset at baseline with posteroanterior view acquired with 10 degrees beam angle (PA10) of the knee given that both joints are graded with KL score.
Knees with total knee replacement were excluded.
Subject wise partitioning was used to split data into train and test sets in a 5-fold cross validation setting.
}
\label{tab: data}
\centering
\begin{tabular}{lcc}
\toprule
 & \textbf{OAI} & \textbf{MOST} \\
\midrule
Number of knees & 9012  & 3644\\
Number of subjects & 4506 & 1822\\
Number of samples where KL\textless{}2 & 5045  & 2247\\
Number of samples where KL\textgreater{}=2 & 3967  & 1397\\
\bottomrule
\end{tabular}
\end{table}

\subsection{Data Preprocessing}

We extracted anatomical landmark points ({keypoints}) of knee image using BoneFinder$^{\textrm{\tiny{\textregistered}}}$ \cite{lindner2013fully} tool.
In the preprocessing pipeline, the 16-bit DICOM images are first normalized using global contrast normalisation and a histogram truncation between the $5^{th}$ and $99^{th}$ percentiles.
These images were eventually converted to 8-bit images ($0-255$ grayscale range).
The image resolution which was not standardized in the database was also standardized to $0.2$ mm using a bicubic interpolation.
Finally, using landmark points, each knee was rotated to horizontally align the tibial plateau.

%===========================================================================
\subsection{Adaptive Region of Interest} 
\label{sec:ROI}

This section describes our developed adaptive region segmentation and ROI selection approach.
Firstly, we segmented tibia and femur from the background using the landmark points. 
Secondly, we performed oversegmentation of the bone region into subregions separately for femur and tibia using superpixel labelling\cite{achanta2012slic} (Figure \ref{fig: adaptive}). 
Superpixel labelling is an oversegmentation strategy for partitioning images into smaller patches that are spatially contiguous and similar in intensity. 
Superpixel clusters produce compact regions (subregions, superpixels) that follow image boundaries.

To generate the superpixels we used Simple Linear Iterative Clustering (SLIC) because of its simplicity and flexibility in the compactness and number of the superpixels it generates\cite{achanta2012slic}.
The algorithm adopts \textit{k-means} clustering in intensity and spatial domain. 
Starting from \textit{k} regularly spaced cluster centers, each pixel in the image is associated with the nearest cluster center. 
The process of associating pixels with the nearest cluster center and recomputing the cluster center  is repeated iteratively until convergence. 
At each iteration, superpixels are reassigned to the average color and position of the associated input pixels.

The distance measure \textit{(D)} in computing the similarity between the pixel 
and the cluster center
is an Euclidean norm in the five-dimensional space (color (CIELAB) + spatial):

\begin{equation}
\label{eq:eq1}
D = {\sqrt{{d_{c}}^{2}+ {\displaystyle\frac{d_{s}}{S}}^{2}m^{2}}},
\end{equation}
where $d_{c}$ is the color distance,
$d_{s}$ is the spatial distance,
\emph{m} is a parameter that weighs the relative importance between the color similarity and the spatial proximity,
and \textit{S} is a parameter indicating the size of the superpixels. 
Compactness, \textit{C}, (i.e.more compact superpixels have lower area to perimeter ratio) can be then controlled by \textit{m}.
The ability to specify the amount of superpixels and the ability to control the compactness are important properties of this method. In addition, SLIC  exhibit state-of-the-art adherence to image boundaries.

\subsubsection*{Detection of the Most Informative Region}

\begin{figure}[htp]
\centering
\resizebox{\textwidth}{!}{
\subfloat[]{\includegraphics[height=4cm,, trim=8mm 0mm 3mmin 0mm,clip]{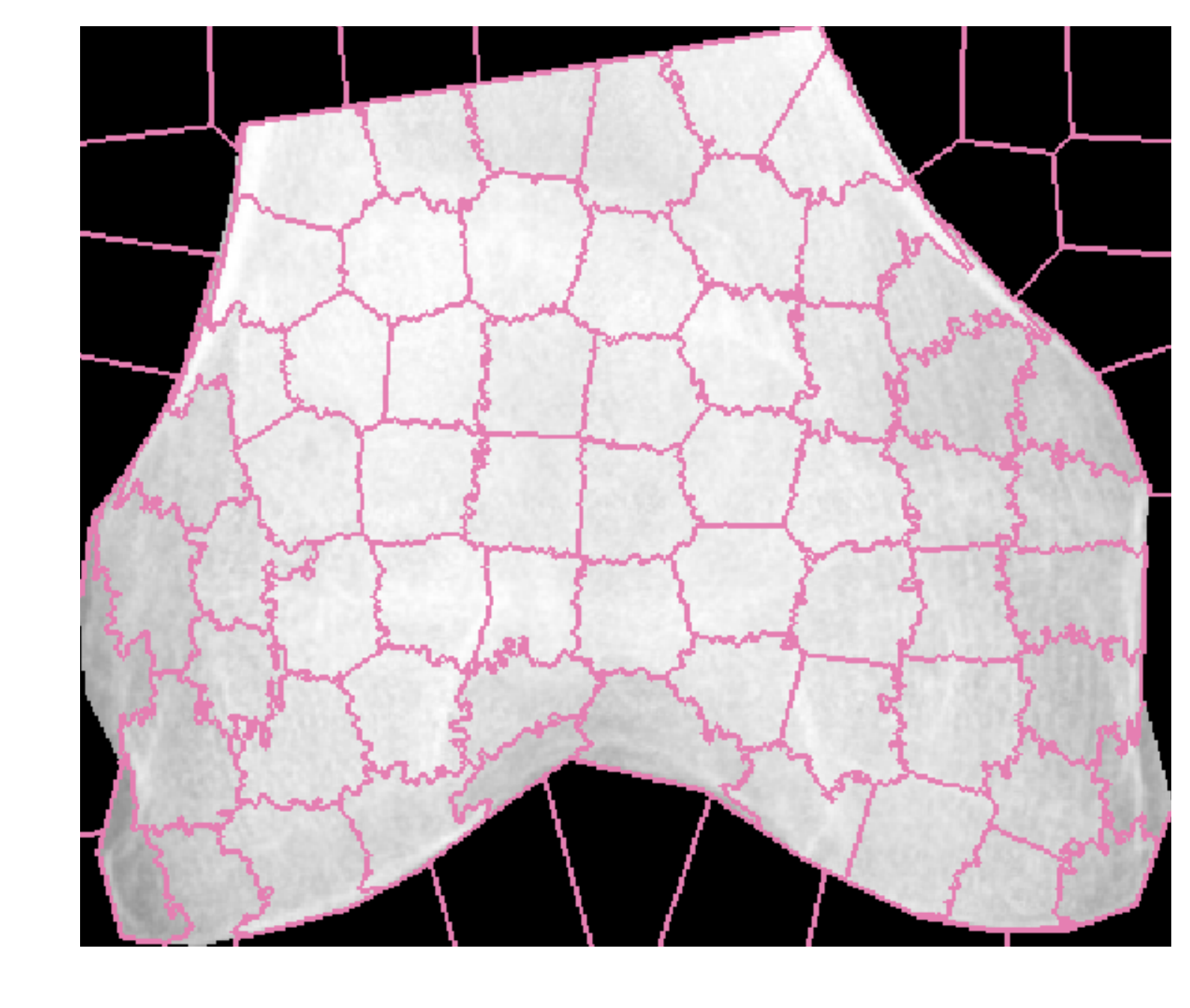}}
\subfloat[]{\includegraphics[height=4cm,, trim=8mm 0mm 3mmin 0mm,clip]{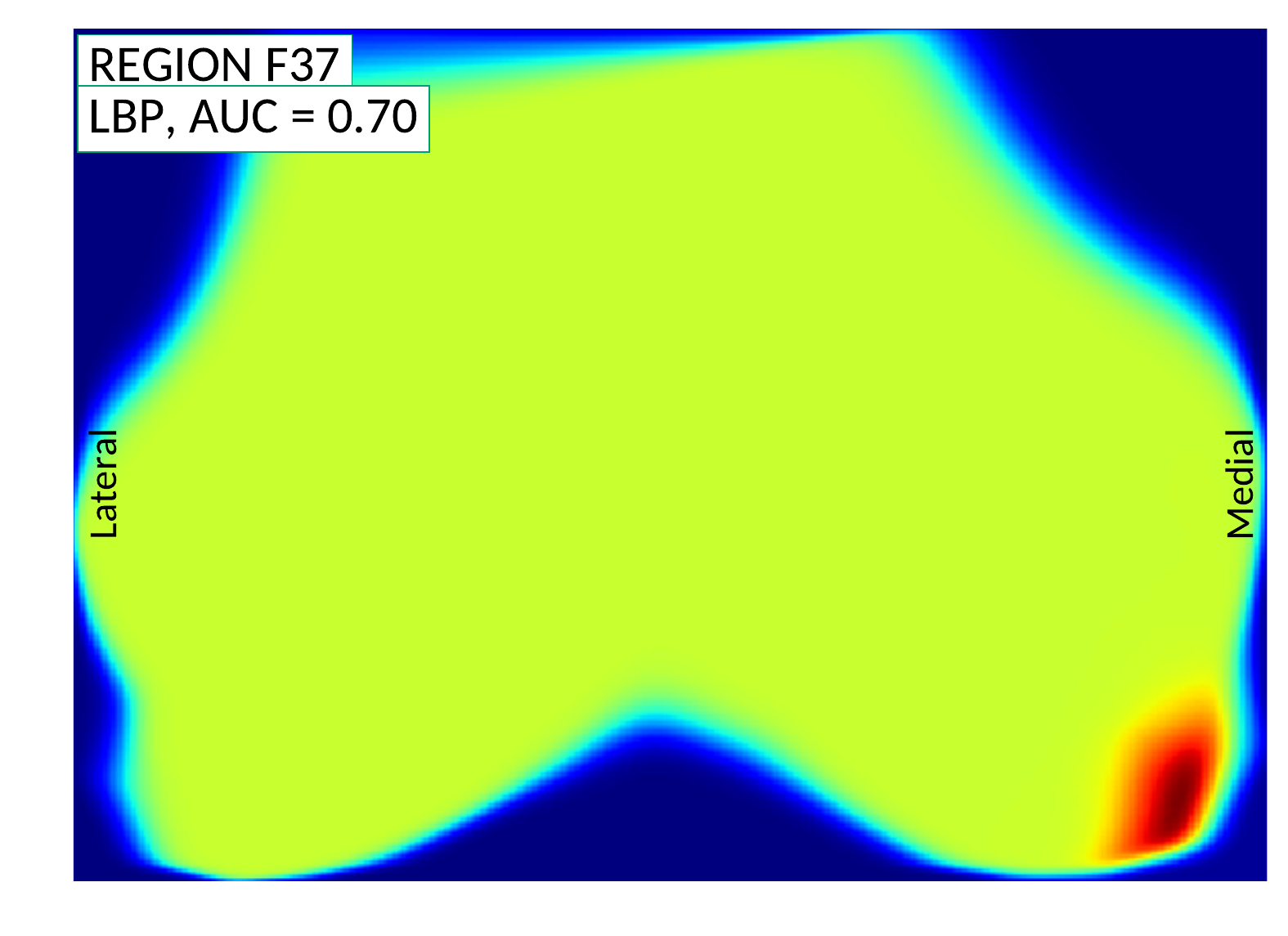}}
\subfloat[]{\includegraphics[height=4cm,, trim=8mm 0mm 3mmin 0mm,clip]{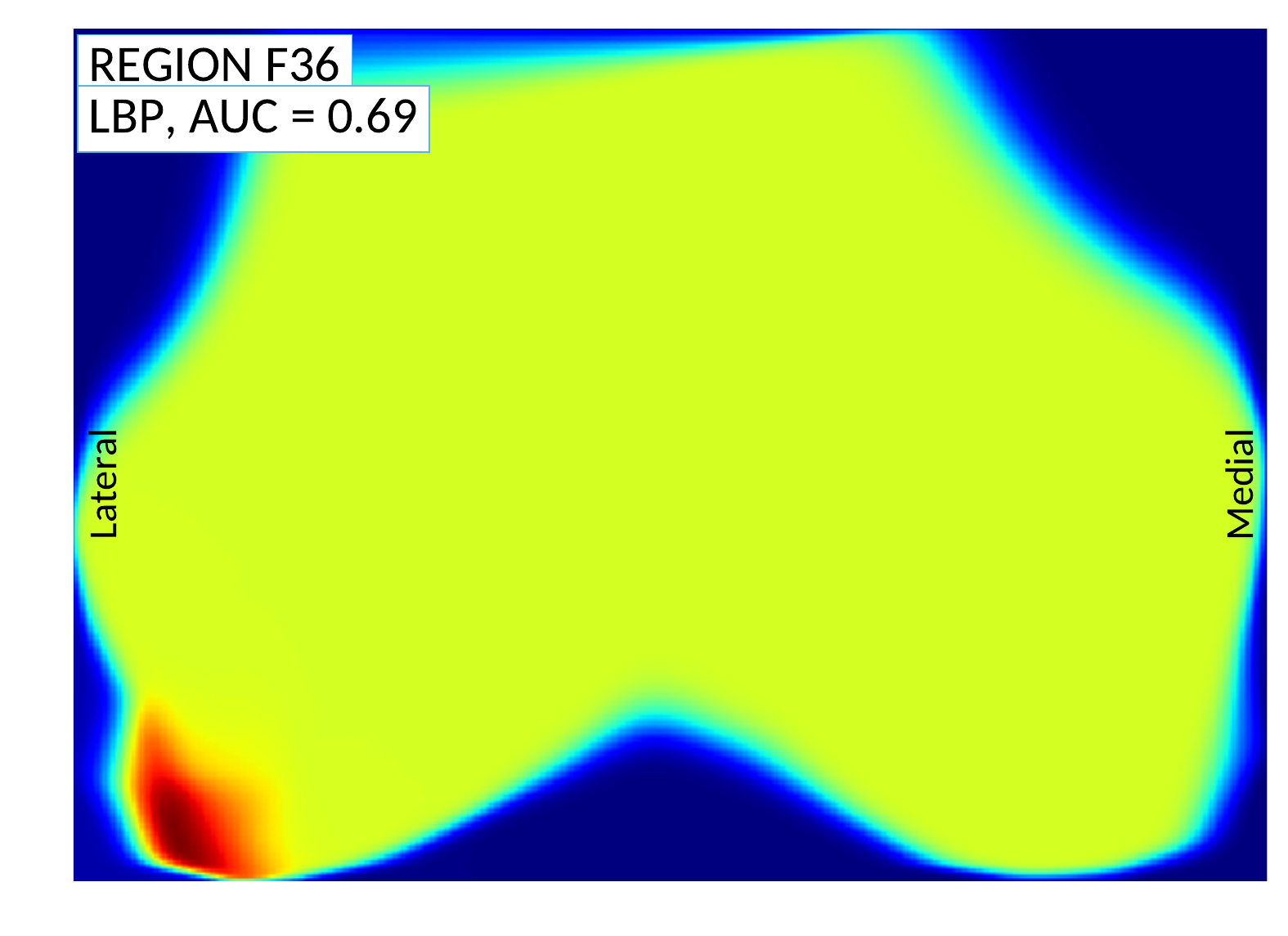}}
}
\resizebox{\textwidth}{!}{
\subfloat[]{\includegraphics[height=4cm, trim=8mm 0mm 3mmin 0mm,clip]{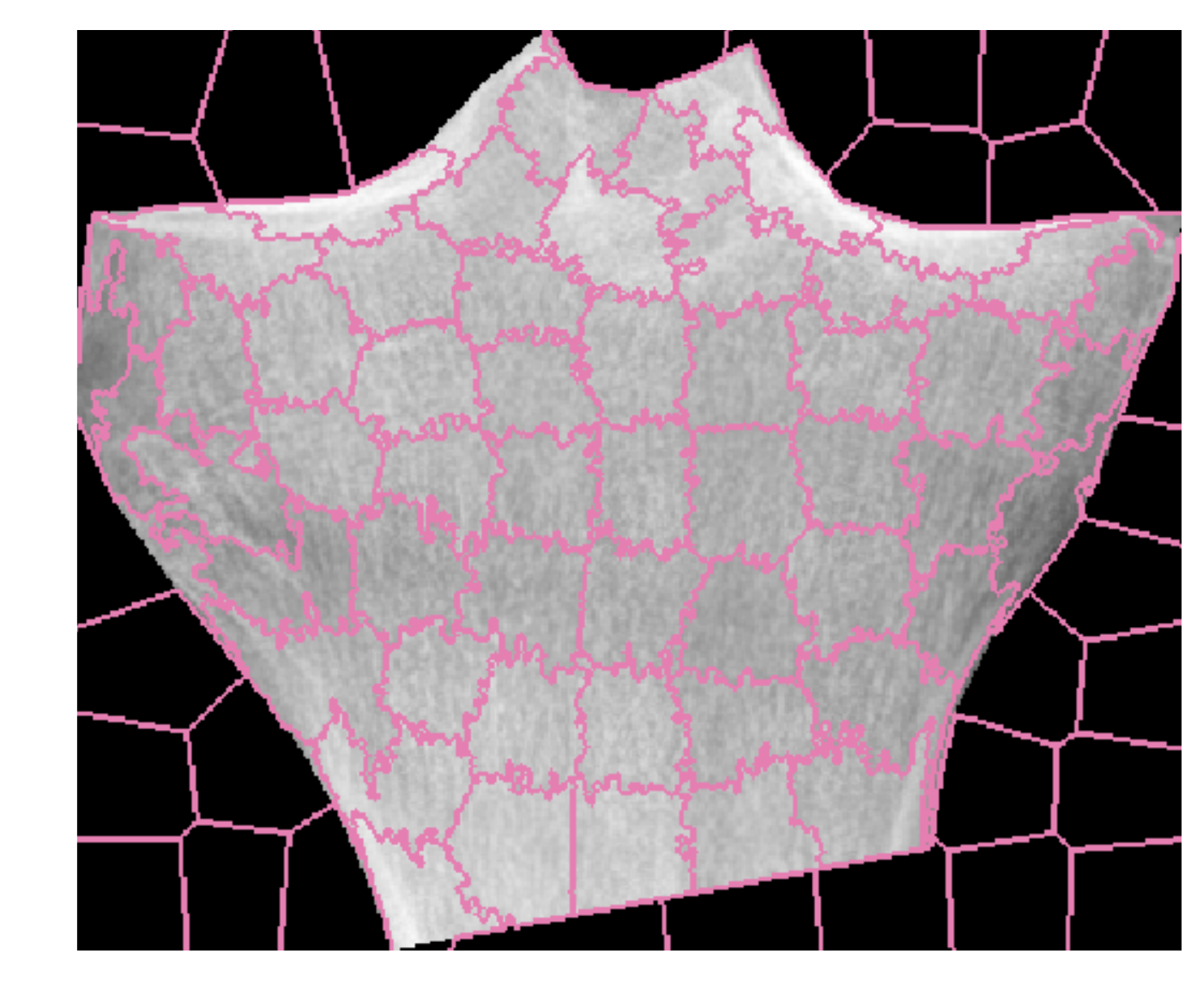}}
\subfloat[]{\includegraphics[height=4cm, trim=8mm 0mm 3mmin 0mm,clip]{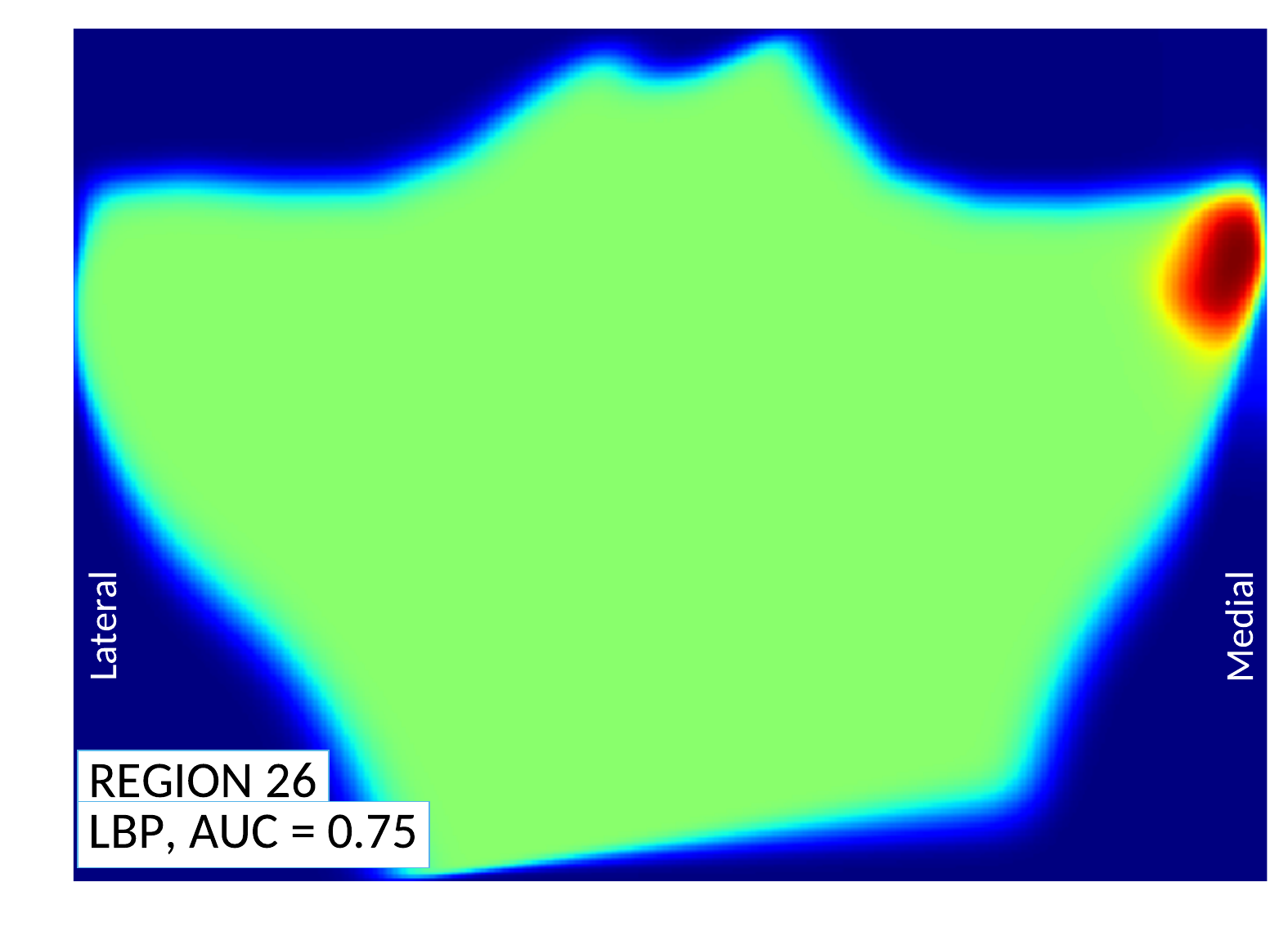}}
\subfloat[]{\includegraphics[height=4cm, trim=8mm 0mm 3mmin 0mm,clip]{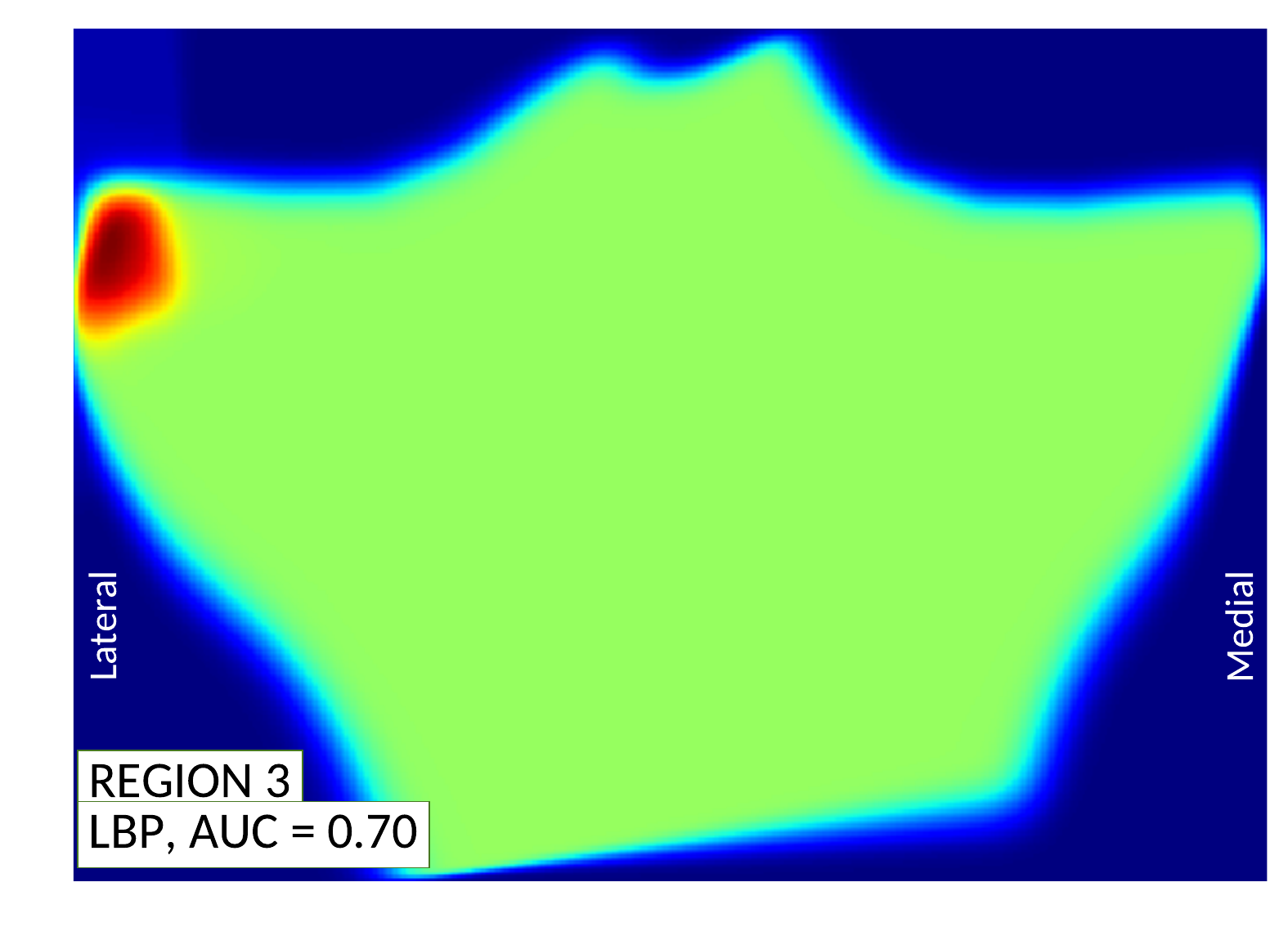}}
}
\caption{
The most informative regions we found are located at the sites where osteophytes are typically occurring.
The subfigure,
(a) demonstrates a sample showing the superpixel segmentation boundaries of a femur image, (b) shows
the most informative region of femur \textcolor{green}{\texttt{f37}} which is located at medial margin, and (c) shows the second most informative region of femur
\textcolor{orange}{\texttt{f36}} which is located at lateral margin,
(d) Demonstrates a sample showing the superpixel segmentation boundaries of a tibia image,
(e) shows the most informative region of tibia \textcolor{magenta}{\texttt{t26}} which is located at medial margin and (f) shows the second most informative region of tibia which is
\textcolor{cyan}{\texttt{t3}}
located at lateral margin based on our grid placement.
}
\label{fig: region}
\end{figure}

\begin{figure}[htp]

\subfloat[Presentation of our adaptive region segmentation and region of interest selection approach.
We first placed a dense grid (57 points) on each subject proportional to its size. 
Subsequently, for each grid point (location), in order to assess its descriptive power we used the segmented region that encloses the grid point.
Segmented region was then described by LBP and the corresponding vector was fed into the machine learning pipeline (see Figure \ref{fig: flowchart}).
Similar analysis was also done for femur compartment.]{\includegraphics[width = 1\linewidth]{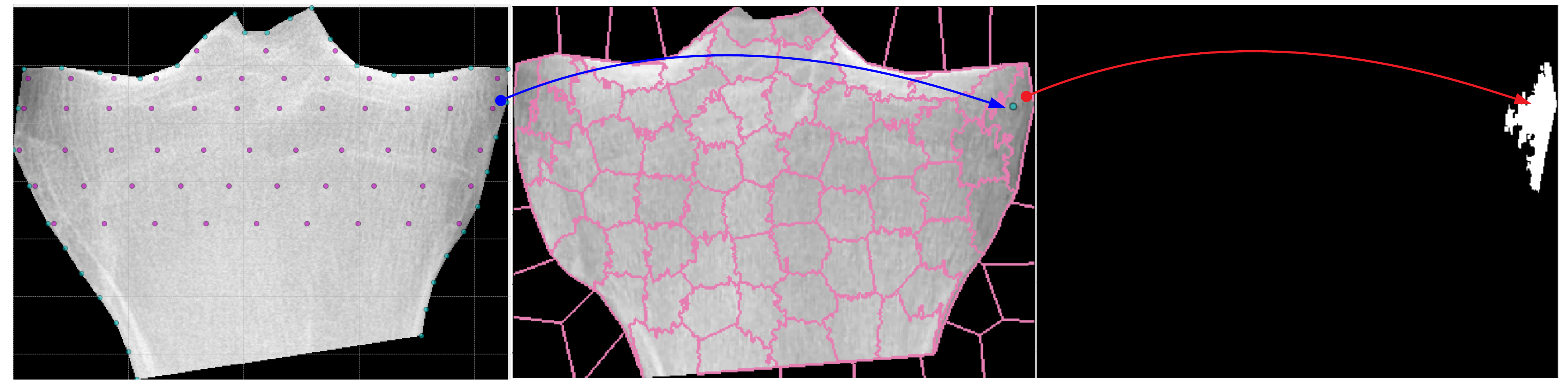} \label{fig: grid}}\\
\subfloat[In order to obtain the mean shape and the location of most informative region, masks of the adaptive regions that corresponds to the same location were accumulated over all subjects.]{\includegraphics[width = 1\linewidth]{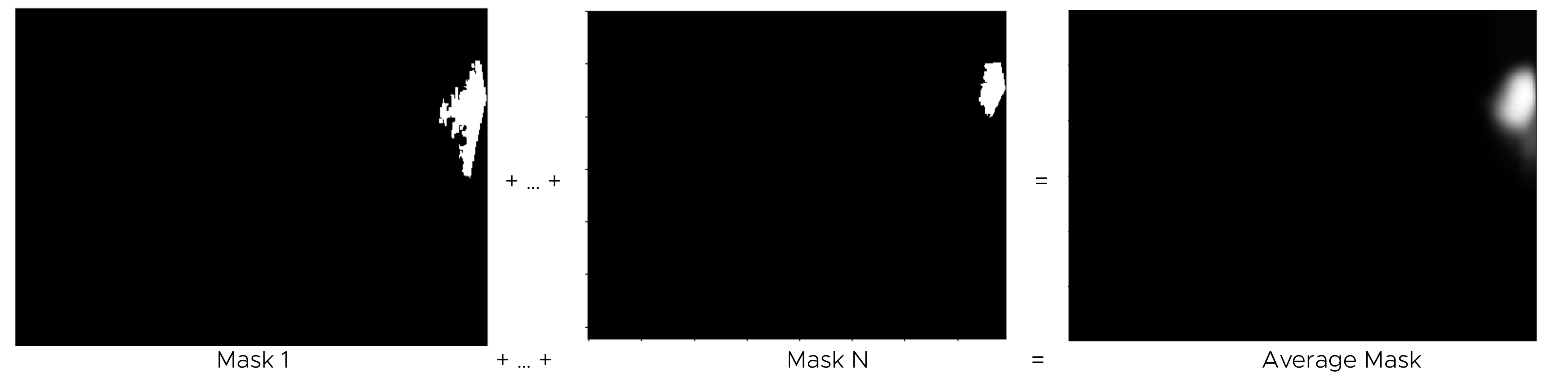} \label{fig: mask}}\\
\subfloat[\textit{(Left to right)} Average mask obtained from region 
\textcolor{magenta}{\texttt{t26}} overlaid on average tibia image. Thresholded mask using Otsu's method (adaptive mask ROI). Boundaries of the adaptive mask ROI is shown on a sample tibia image.]{\includegraphics[width = 1\linewidth]{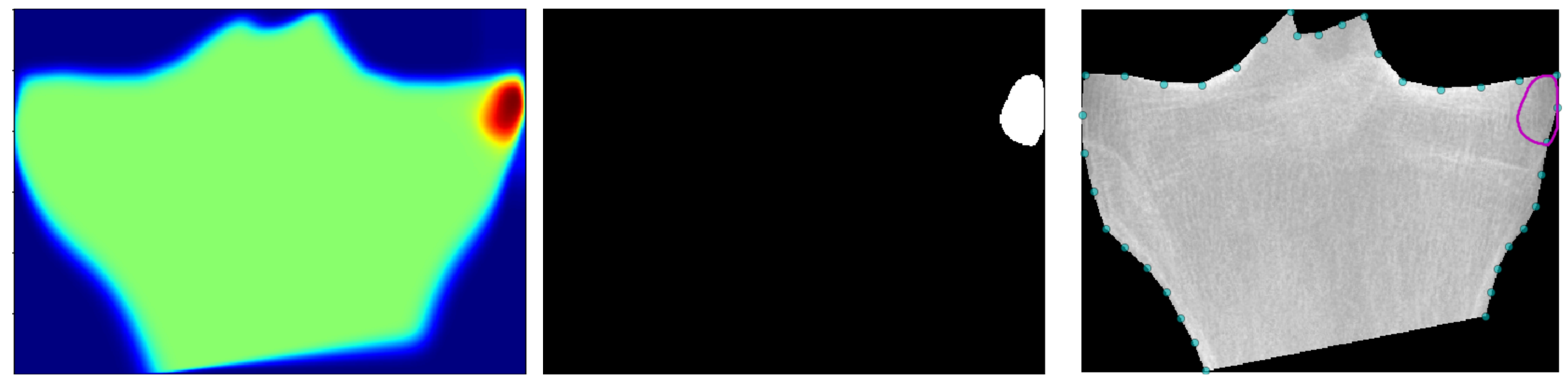}\label{fig: region26}}
\caption{Overview of the developed adaptive region segmentation and the region-of-interest (ROI) selection approach.}
\label{fig: adaptive}
\end{figure}

In the following step, we used $M$ regularly placed grid points, $g_i, i \in {1,2,..,.M}$, in order to select approximately the same subregion for all subjects (Figure \ref{fig: grid}) for feature extraction and machine learning stages.
Grid points do not necessarily end up in different regions. 
We intentionally placed relatively dense grid compared to the average size of superpixels to cover the full bone region.
Therefore, multiple points may fall in the same subregion (superpixel) but this does not hamper the analysis.  
Each subregion $r_i$ is then described by a feature vector $f\{r_i\}$.
Finally, we evaluated and compared $M$ regions based on textural properties.
In this step, we employed Local Binary Patterns (LBP) descriptor to detect the most informative (optimal) region.
Here, we defined the most informative region as the subregion where the texture classifier (based on LBP features) performs best to distinguish OA samples from non-OA.
Figure \ref{fig: region} shows the first two informative regions on tibia and femur based on grid locations.
We provided the full analysis in the \nameref{sec:supp} material.
Instead of utilizing grid based locations, we made a second pass to detect the adaptive regions that correspond to explicit pixel locations which fall (2mm, 2mm) inside the medial tibia margin point and the lateral tibia margin point and utilized them in the rest of the experiments.
Heat-map representations of these regions were obtained by averaging the corresponding subregions over all subjects in the database (average mask).
The results suggest that outer medial tibia side provides richer information than other regions from the point of texture analysis.

Figure \ref{fig: mask} shows the process of averaging and also individual masks that correspond to the outermost medial region which is denoted by \textcolor{magenta}{\texttt{t26}} (tibia region 26) that follows the lattice numbering.
The average mask for the most informative region shows that there is a large portion of overlap between the segmentation masks.
We take the advantage of this finding to reduce the computational complexity of the overall method in the following experiments.
Instead of segmenting the tibia of each individual subject, we employed the mask which was obtained by thresholding the average mask using Otsu's method.
Thresholded mask and its contour on a sample image are presented in Figure \ref{fig: region26}.

%========================================================================================

\subsection{Standard ROI}

\begin{figure}[!htb]
\centering
\subfloat[]{\includegraphics[height = 6cm]{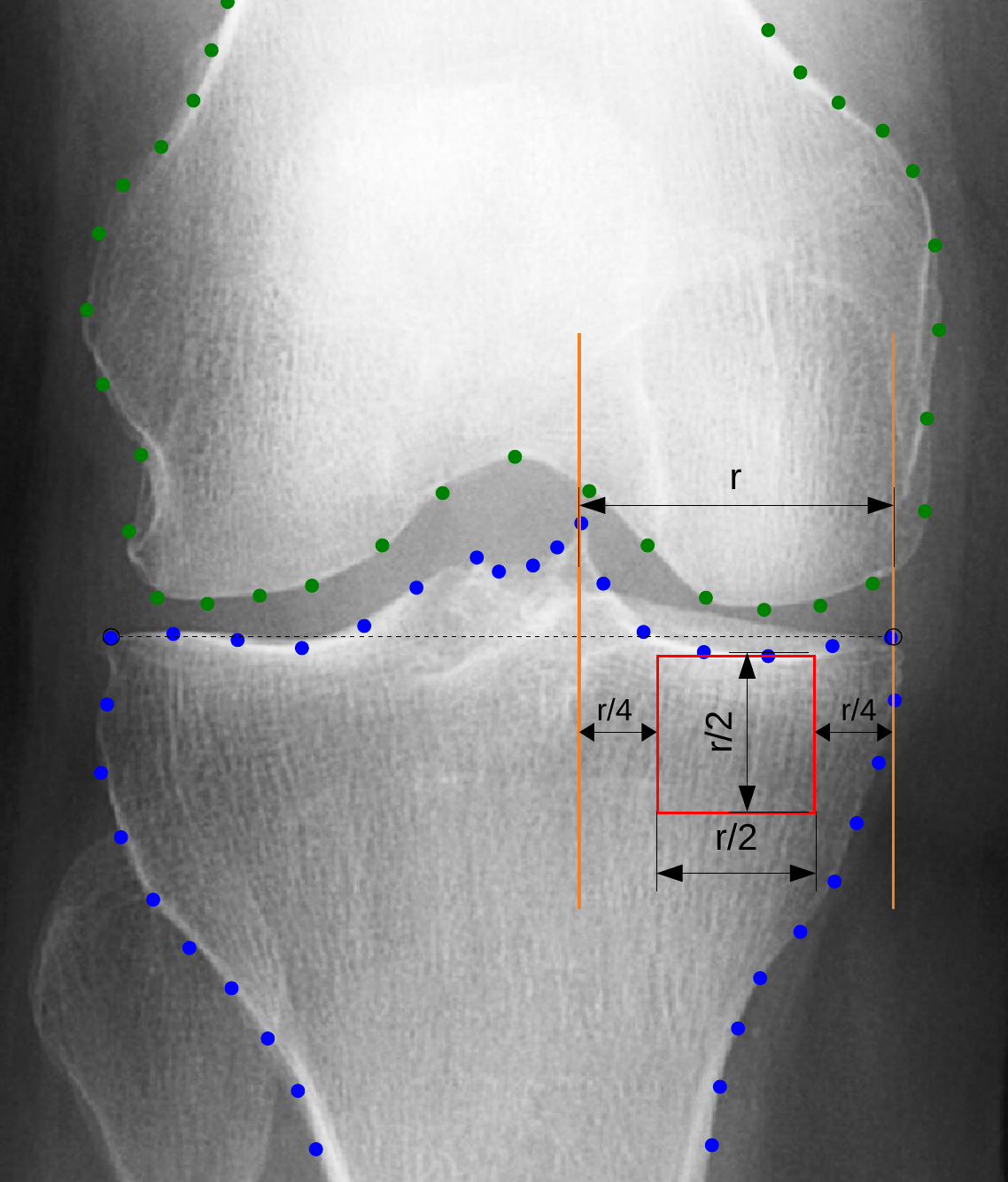}}
\qquad
\subfloat[]{\includegraphics[]{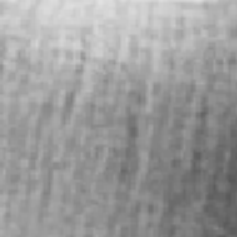}}
\caption{Illustration of (a) the placement of standard region of interest (ROI) and (b) corresponding (cropped) ROI image.}
\label{fig: stdroi}
\end{figure}

We utilized the extracted landmark points to locate standard rectangular ROI in a fixed region on each knee.
We followed the literature \cite{lynch1991analysis,hirvasniemi2017differences,wolski2011trabecular,thomson2015automated,hafezi2018new} 
to locate `standard' ROI: 
We used a square patch placed immediately beneath the tibial plateau with dimensions proportional to the width of the knee  in the centre of the
medial condyle of tibia (see Figure \ref{fig: stdroi}) .

%=========================================================================================

\subsection{Analysis of Texture Features}

\subsubsection*{Texture Descriptors}
Although texture descriptors have been applied to plain knee radiographs for years,
Fractal Signature Analysis (FSA) or fractal dimension (FD) and its variations remain the main method to analyse OA texture until today
\cite{lynch1991analysis, janvier2017subchondral,hirvasniemi2017differences,buckland2004subchondral,ljuhar2018combining,jennane2014variational,kraus2018predictive,hirvasniemi2014quantification,wolski2010differences,thomson2015automated,podsiadlo2016baseline,messent2005tibial,kraus2009trabecular, wolski2009directional,janvier2015roi, thomson2015automated}
since 1990's.
Other texture descriptors used in OA analysis include 
simple pixel features \cite{thomson2015automated},
Haralick's texture features based on the Gray Level Co-occurrence Matrix (GLCM)\cite{shamir2009knee},
Gabor filter banks \cite{anifah2013osteoarthritis}, 
Entropy \cite{hladuuvka2017femoral}, 
Wavelet Transform \cite{riad2018texture},
Tamura texture features\cite{shamir2009knee},
and Local Binary Patterns (LBP) \cite{hirvasniemi2014quantification}.
Sometimes the combination of different texture descriptors and also shape descriptors have been used \cite{shamir2009early,shamir2009knee,thomson2015automated}.

Fractals, which is a measure of surface `roughness', have been used to analyze and quantify very complex shapes or structures.
FD is calculated by taking the pixel intensity differences at varying scales of the image.
Then the slope of the line which is fitted to a log-log plot of the intensities across the scales, determines the FD.
In OA related studies, it was first used by Lynch et al. \cite{lynch1991analysis} to quantify trabecular bone texture. 
The FD was linked to spacing, variation in thickness and orientation, and cross-connectivity of the trabeculae structure \cite{kraus2009trabecular}.
However, FSA is sensitive to image artifacts and noise \cite{veenland1996estimation}.

As noted earlier, a variety of methods for FD calculation has been proposed in the literature (power spectrum method, maximum likelihood method, tile counting method, box counting method, blanket methods, modified Hurst orientation transform, variance orientation transform, augmented variance orientated
transform).
Differences in the techniques for measuring FD result different numerical values, albeit they aim at estimating surface {roughness}.
Although they are correlated, they do not directly measure the same image property.
Therefore, it is not easy to reproduce and compare fractal based approaches.
In addition, the narrow range of the FD measurement values susceptible to limited discrimination power.

GLCM represents the distance and angular spatial relationship of pixels which is then used to derive several statistics.
Haralick proposed a set of fourteen feature measures based on GLCM including 
contrast,
correlation, entropy, variance, sum average, sum variance,
sum entropy, angular second moment, 
difference variance, difference entropy, information
measure of correlation 1, information measure of
correlation 2, and inverse difference moment.
The well-known local binary descriptor “Local Binary Patterns” labels image pixels by thresholding
the neighbourhood of each pixel with the center value and considering the
result as a binary number \cite{ojala2000gray}.
Then distribution of LBP code of an image is used to describe the texture by a histogram vector.
Shannon entropy which measures the amount of randomness of gray levels is also used to characterize the texture of an input image. 
Histogram of Oriented gradients (HOG)\cite{dalal2005histograms} employs distribution of the directions of the image gradients. 
Image is divided into a overlapping rectangular grid of cells grouped into $k\times k$ blocks. 
A histogram of the gradient directions is then computed within each cell of each block. 

\subsubsection*{Implementation Details}

After prepossessing the raw radiography data, we select ROI for computing the texture descriptors.
We utilized both classical approach (rectangular ROI) and adaptive ROI in our experiments.
We evaluated and compared FD, LBP, Haralick features, Shannon entropy, and HOG.

For FD computation, we used the implementation from \cite{hirvasniemi2016correlation} which calculates FD separately for vertical and horizontal trabecular structures within the ROI.
Compared to \cite{hirvasniemi2016correlation}, we increased the maximum size of the flat disk from 2mm to 3.2mm which is used as a structuring element for calculating the FD.
We also noticed that the size of the structuring element affects the performance of the method such that bigger structuring element results slightly better.
We utilized only the first 13 features from Haralick descriptor using publicly available open source Mahotas library \cite{coelho2012mahotas}.
For other descriptors and SLIC, we used scikit-learn \cite{scikit-learn} python package which is also an open source library.
For all experiments, we chose empirically the compactness parameter as 0.08 and the number of regions as 100 for the initial region segmentation (i.e. SLIC parameters).
The best parameters for standard ROI for LBP and HOG were determined by grid search technique in scikit-learn \cite{scikit-learn} package.
In order to favor the standard ROI, the optimal parameters on standard ROI were employed in adaptive ROI experiments.
Search space of parameters and the best values that optimized the AUC value on cross-validation are given in \nameref{sec:supp} material in Table \ref{tab:hyperparameter}.

\subsection{Statistical Analysis}

In order to assess how the results of our analysis generalize over a set of independent data we used i) cross-validation and ii) validated our trained model with an independent test set that is completely different from the training one.
Cross-validation is a method for estimating predictive performance and also for model selection. 
For large sample sizes, the K-fold cross-validation method gives nearly unbiased estimator.
In K-fold cross-validation, the training data is divided into $K$ disjointed parts of approximately equal size.
Then the learning algorithm is trained on $K-1$ of $K$ independent subsets (training set) and the remaining one (test set) is used to estimate the predictive performance  (expected loss on unseen future samples).
The algorithm is trained $K$ times, each time using a different test partition.
The estimate of error rate is the average of the errors incurred on all folds. 
Here we evaluate subject-wise cross-validation which is more reliable than  record-wise cross-validation to assess the prediction accuracy of a machine learning algorithm.
The error rate measures how well the two classes in the data set are separated.
We present the area under the receiver operating characteristic curves (ROC AUC) which is also called $c-index$, a common metric to measure classifier performance effectively.
ROC AUC provides is a combined measure of sensitivity and specificity.
The higher the AUC, the better the predictive performance of the classification method.
In addition, we present average precision score which quantifies precision-recall curves similar to AUC. 
The implementation is done using scikit-learn \cite{scikit-learn} package.

We used two-class regularized logistic regression (LR) to predict the image level label (OA vs non-OA).
To mitigate overfitting, we employed regularization. 
Features obtained by texture descriptors were properly standardized using mean and variance of the training data of each fold.

%% file: sections/results_conclusion.tex
\section{Results}
\input{sections/tables.tex}

\section{Discussion}
In this paper, we investigated the placement of ROI for SB texture analysis of knee radiographs. Using automatic oversegmentation method, we found that the most informative bone region associated in OA was located at the tibial margin in the medial side of the knee. We observed that this ROI has a profound effect on bone texture analysis, and this could be due to absorption of uneven mechanical load across the joint \cite{egloff2012biomechanics} and osteophytes that form along joint margins.

The framework presented in this study demonstrated that the performance of the current state-of-the-art approaches used for texture analysis could significantly be improved using the proposed adaptive ROI approach.
However, results regarding the entropy and FD are mixed. This could be explained by the higher sensitivity of entropy and FD to the noise and image size \cite{kadir2001saliency,veenland1996estimation}.

We observed that LBP yielded the best performance in all experimental settings with AUC of 0.761 [0.751, 0.771] and AP of 0.737 [0.724, 0.749] on OAI, AUC of 0.797 [0.781, 0.812] and AP of 0.754 [0.735, 0.772] on MOST, and AUC of 0.818 [0.802, 0.832] and AP of 0.779 [0.760, 0.795] on validation setting when evaluated on adaptive mask ROI (\hyperref[fig: region26]{\textcolor{magenta}{\texttt{t26}}}).
HOG showed the second best performance and yielded slightly lower AUC and AP compared to LBP.
We obtained similar classification performances with FD and Haralick features. 
We also observed that entropy lacks discriminative power in classification of bone texture patches.
Although both entropy and FD are widely adopted texture descriptors in OA research \cite{lynch1991analysis, janvier2017subchondral,hirvasniemi2017differences,buckland2004subchondral,ljuhar2018combining,jennane2014variational,kraus2018predictive,hirvasniemi2014quantification,wolski2010differences,thomson2015automated,podsiadlo2016baseline,messent2005tibial,kraus2009trabecular, wolski2009directional,janvier2015roi,thomson2015automated,hladuuvka2017femoral}, they performed significantly poorer than LBP in our experiments.

When we used the test data (MOST) which is independent from the training data (OAI) in our validation experiments (Exp 3), we observed that classification performance was not affected, even slight improvements were obtained with LBP and HOG descriptors.
This may be explained by different distribution and amount of of training samples (2915 vs 9012),
and, potentially, imperfect annotations of KL grades \cite{tiulpin2018automatic, norman2019applying}.

Combination of different texture features (e.g. HOG and LBP) did not provide a significant performance improvement over a single descriptor.
For some cases, the classification performance even decreased, which can be explained by multicollinearity, i.e., several individual texture features are strongly inter-correlated leading to numerically unstable regression models. On the other hand, when we concatenated a particular texture descriptor from medial and lateral side, the classification performance improved (Table \ref{tab:combine_region}).
This is an indication that lateral and medial margin may provide complementary information regarding the bone textural changes in OA.

We observed that, in OA, the radiographically most distinctive bony changes occur at the medial tibia margin, and all the compartments including subchondral cortical plate and subchondral trabecular bone are affected.
The alignment of the adaptive region could be correlated with the uneven force that tibia experiences in OA \cite{egloff2012biomechanics}.
It could also follow the knee alignment due to deformity.
Therefore, it would be interesting to investigate the associations between adaptive ROI alignment and mechanical/anatomic axis angle.

In the light of the experimental results, we believe that the performance of texture analysis to quantify bony changes in radiographic OA could be improved using more robust texture descriptors than the most popular FD. The experimental results of the texture descriptors like LBP and HOG for detecting the radiographic OA presence are convincing, and these approaches are not that sensitive for changes in radiographic acquisition protocols\cite{janvier2017subchondral}. Thus, they could be applied in clinical decision support tools in the future. In addition to better texture descriptors, the predictive ability of such automated tools could benefit from inclusion of other image features, such as joint shape, osteophytes and joint space width.

Although we have presented a novel method to localize the optimal ROI to improve the performance of the texture analysis, this study has still some limitations.
First, the general limitation in application of the machine learning methods to a clinical setting is generalizability of the model due to potential bias associated with the training data and inherited bias due to algorithms. 
However, we validated the trained model with an independent test set that is completely different from the training one. 
Furthermore, in order to assess the stability of the method we used cross-validation setting.
Second, texture descriptors could be sensitive to imaging conditions such as rotations, beam angle, noise, exposure, blur, accelerating voltage, and image post-processing on digital X-ray systems.
However, we tried to address this limitation by pre-processing the image data, yet
the results may still be affected. 
Third, we selected the size and compactness parameters of the superpixel segmentation method empirically.
Therefore, further investigations are required to clarify the effect of parameters used in adaptive segmentation stage. 
Moreover, we utilized the average mask obtained from the most informative region to reduce the computational complexity, which could lower the performance.
Fourth, we believe that the performance of landmark detection algorithm has a direct effect on the analysis which is another limitation of our method. In this study, we relied on BoneFinder$^{\textrm{\tiny{\textregistered}}}$ \cite{lindner2013fully} tool for landmark detection, which could be improved \cite{tiulpin2019kneel}. 
Finally, although the most informative regions we found are 
located at the sites where osteophytes are typically occur, we did not analyze the direct effect of the presence of osteophytes and their surrounding.
More research is needed in order to assess the role of osteophytes in the SB texture, specifically within the ROIs located at bone margins.

To conclude, we believe that texture analysis methods have the potential to reveal structural changes in SB, and in particular, the association between early OA and remodeling of the fine trabecular network. 
This study demonstrated and confirmed that the localization of ROI plays a significant role in bone texture analysis.
Our findings show that placing the ROIs at tibial margins could provide more discriminative information OA changes in the bone and lead to more sensitive imaging-based biomarkers.

\section*{\small{Acknowledgments}}
The OAI is a public-private partnership comprised of five contracts (N01-AR-2-2258; N01-AR-2-2259; N01-AR-2-2260; N01-AR-2-2261; N01-AR-2-2262)
funded by the National Institutes of Health, a branch of the Department of Health and Human Services, and conducted by the OAI Study Investigators. 
Private funding partners include Merck Research Laboratories; Novartis Pharmaceuticals Corporation, GlaxoSmithKline; and Pfizer, Inc. Private sector funding for the OAI is managed by the Foundation for the National Institutes of Health. This manuscript was prepared using an OAI public use data set and does not necessarily reflect the opinions or views of the OAI investigators, the NIH, or the private funding partners

Multicenter Osteoarthritis Study (MOST) Funding Acknowledgment. MOST is comprised of four cooperative grants (Felson – AG18820; Torner – AG18832, Lewis – AG18947, and Nevitt – AG19069) funded by the National Institutes of Health, a branch
of the Department of Health and Human Services, and conducted by MOST study investigators. This manuscript was prepared using MOST data and does not necessarily reflect the opinions or views of MOST investigators.

We would like to acknowledge the strategic funding of the University of Oulu, Infotech Oulu.

Dr. Claudia Lindner is acknowledged for providing BoneFinder.

\section*{\small{Author contributions}}

N.B. originated the idea.
N.B., A.T., and S.S. designed the study.
J.H. and M.T.N. participated in conceptualization of the study.
N.B. and A.T. conducted the experiments.
S.S. contributed at each stage of the process and supervised the project.
All authors contributed to interpreting the data, writing and editing the manuscript, and have approved the submitted version of the manuscript.

\section*{\small{Role of the funding source}}
Funding sources are not associated with the scientific contents of the study.

\section*{\small{Conflict of interest}}
The authors report no conflicts of interest.

%% file: sections/tables.tex
\begin{table}
 \caption{\small{Comparison of texture descriptors on OAI calculated from Standard ROI vs Adaptive ROI. Scores are given by ROC AUC and Precision-Recall AP with 95\% confidence interval in parentheses on a 5-fold cross validation setting.}}%
  \label{tab:OAI_region_comparison}%
  \centering
  \begin{tabular}{lccc}
    \toprule
    & & \multicolumn{2}{c}{\textbf{OAI - \hypertarget{exp1}{Exp 1}} }\\
    Method      & Score     & Standard ROI                & Adaptive mask \hyperref[fig: region26]{\textcolor{magenta}{\texttt{t26}}} \\
    \midrule
    LBP           & AUC & 0.685 [0.674, 0.696] & \textBf{0.761 [0.751, 0.771]}\\
    & AP  & 0.653 [0.639, 0.665] & \textBf{0.737 [0.724, 0.749]}\\
    Fractal & AUC & 0.621 [0.609, 0.633] & \textBf{0.664 [0.652, 0.675} \\
    & AP  & 0.570 [0.556, 0.584] & \textBf{0.611 [0.597, 0.624]}\\
    HOG           & AUC & 0.651 [0.640, 0.662] &\textBf{0.742 [0.731, 0.752]}\\
     & AP  & 0.603 [0.590, 0.616] & \textBf{0.708 [0.695, 0.720]}\\
     Haralick      & AUC & 0.618 [0.607, 0.629] & \textBf{0.667 [0.655, 0.678]} \\
     & AP  & 0.573 [0.560, 0.585] & \textBf{0.634 [0.620, 0.647]}\\
     Entropy       & AUC & 0.585 [0.573, 0.596] & \textBf{0.587 [0.574, 0.598]}\\
     & AP  & \textBf{0.543 [0.529, 0.555]} & 0.530 [0.517, 0.543]\\
    \bottomrule
  \end{tabular}
\end{table}

\begin{table}
 \caption{\small{Comparison of texture descriptors on MOST calculated from Standard ROI vs Adaptive ROI. Scores are given by ROC AUC and Precision-Recall AP with 95\% confidence interval in parentheses  on a 5-fold cross validation setting.}}%
  \label{tab:most_region_comparison}%
  \centering
  \begin{tabular}{lccc}
    \toprule
    & & \multicolumn{2}{c}{\textbf{MOST - \hypertarget{exp2}{Exp 2}} }\\
    Method      & Score     & Standard ROI                & Adaptive mask \hyperref[fig: region26]{\textcolor{magenta}{\texttt{t26}}} \\
    \midrule
    LBP           & AUC &0.778 [0.760, 0.792] & \textBf{0.797 [0.781, 0.812]}\\
    & AP  & 0.733 [0.712, 0.751] & \textBf{0.754 [0.735, 0.772]}\\
    Fractal & AUC & 0.704 [0.685, 0.720] & \textBf{0.729 [0.711, 0.745]} \\
    & AP  & 0.600 [0.576, 0.624] & \textBf{0.652 [0.628, 0.673]}\\
    HOG           & AUC & 0.738 [0.721, 0.754] & \textBf{0.790 [0.774, 0.804]}\\
     & AP  & 0.682 [0.659, 0.702] & \textBf{0.704 [0.678, 0.725]}\\
     Haralick     & AUC & 0.701 [0.682, 0.717] & \textBf{0.791 [0.774, 0.806]} \\
     & AP  & 0.589 [0.565, 0.612] & \textBf{0.744 [0.724, 0.763]}\\
     Entropy       & AUC & \textBf{0.672 [0.653, 0.688]} & 0.610 [0.591, 0.628]\\
     & AP  & \textBf{0.569 [0.545, 0.591]} & 0.516 [0.493, 0.537]\\
    \bottomrule
  \end{tabular}
\end{table}

\begin{table}
 \caption{\small{Comparison of texture descriptors, on an independent test set (MOST) where training data is OAI, calculated from Standard ROI vs Adaptive ROI. Scores are given by ROC AUC and Precision-Recall AP with 95\% confidence interval in parentheses  on a 5-fold cross validation setting.}}%
  \label{tab:independent_region_comparison}%
  \centering
  \begin{tabular}{lccc}
    \toprule
    & & \multicolumn{2}{c}{\textbf{Train: OAI, Test: MOST - \hypertarget{exp3}{Exp 3}} }\\
    Method      & Score     & Standard ROI                & Adaptive mask \hyperref[fig: region26]{\textcolor{magenta}{\texttt{t26}}} \\
    \midrule
    LBP           & AUC & 0.778 [0.761, 0.794] & \textBf{0.818 [0.802, 0.832]}\\
    & AP  & 0.717 [0.694, 0.737] &\textBf{0.779 [0.760, 0.795]}\\
    Fractal & AUC & \textBf{0.699 [0.680, 0.715]} & 0.682 [0.663, 0.699] \\
    & AP  & \textBf{0.602 [0.577, 0.624}] & 0.582 [0.558, 0.606]\\
    HOG           & AUC & 0.718 [0.700, 0.734] & \textBf{0.800 [0.785, 0.813]} \\
     & AP  & 0.640 [0.617, 0.661] & \textBf{0.725 [0.703, 0.746]} \\
     Haralick      & AUC & 0.683 [0.664, 0.700] & \textBf{0.762 [0.745, 0.777]} \\
     & AP  & 0.585 [0.561, 0.607] & \textBf{0.711 [0.689, 0.729]} \\
     Entropy    & AUC   & \textBf{0.673 [0.654, 0.690]} & 0.611 [0.591, 0.630]\\
     & AP  & \textBf{0.571 [0.547, 0.593]} & 0.517 [0.493, 0.538] \\
    \bottomrule
  \end{tabular}
\end{table}

We compared the performance of texture descriptors on the standard ROI and on the best adaptive mask ROI (\hyperref[fig: region26]{\textcolor{magenta}{\texttt{t26}}}).
Tables \ref{tab:OAI_region_comparison},\ref{tab:most_region_comparison}, and \ref{tab:independent_region_comparison} show the performance of Logistic Regression classification for five different texture descriptors with different test and training data.

\subsubsection*{Cross Validation}
\begin{figure}[!htb]
\centering
\subfloat[]{\includegraphics[width = 1\linewidth]{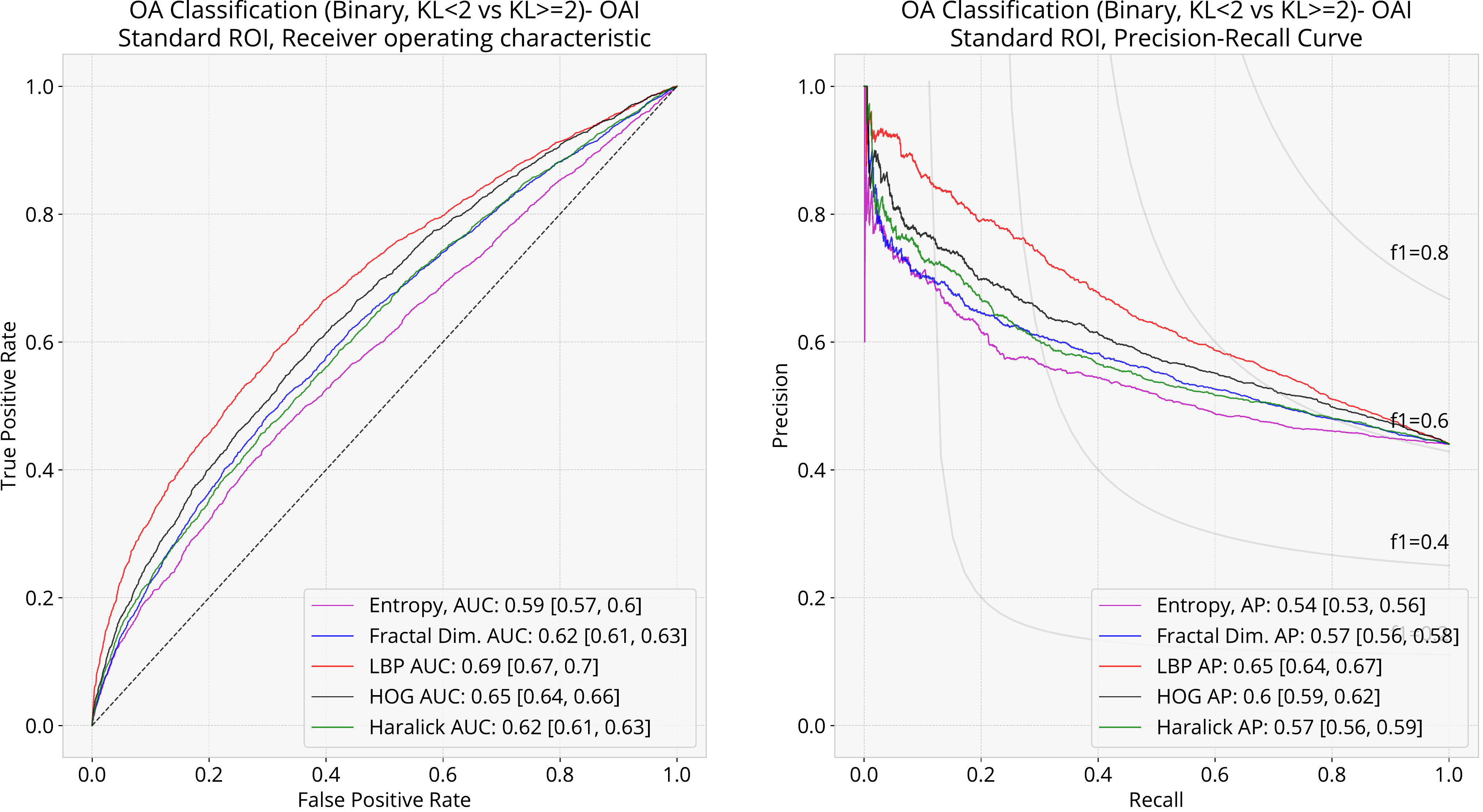}}
\hfill
\subfloat[]{\includegraphics[width = 1\linewidth]{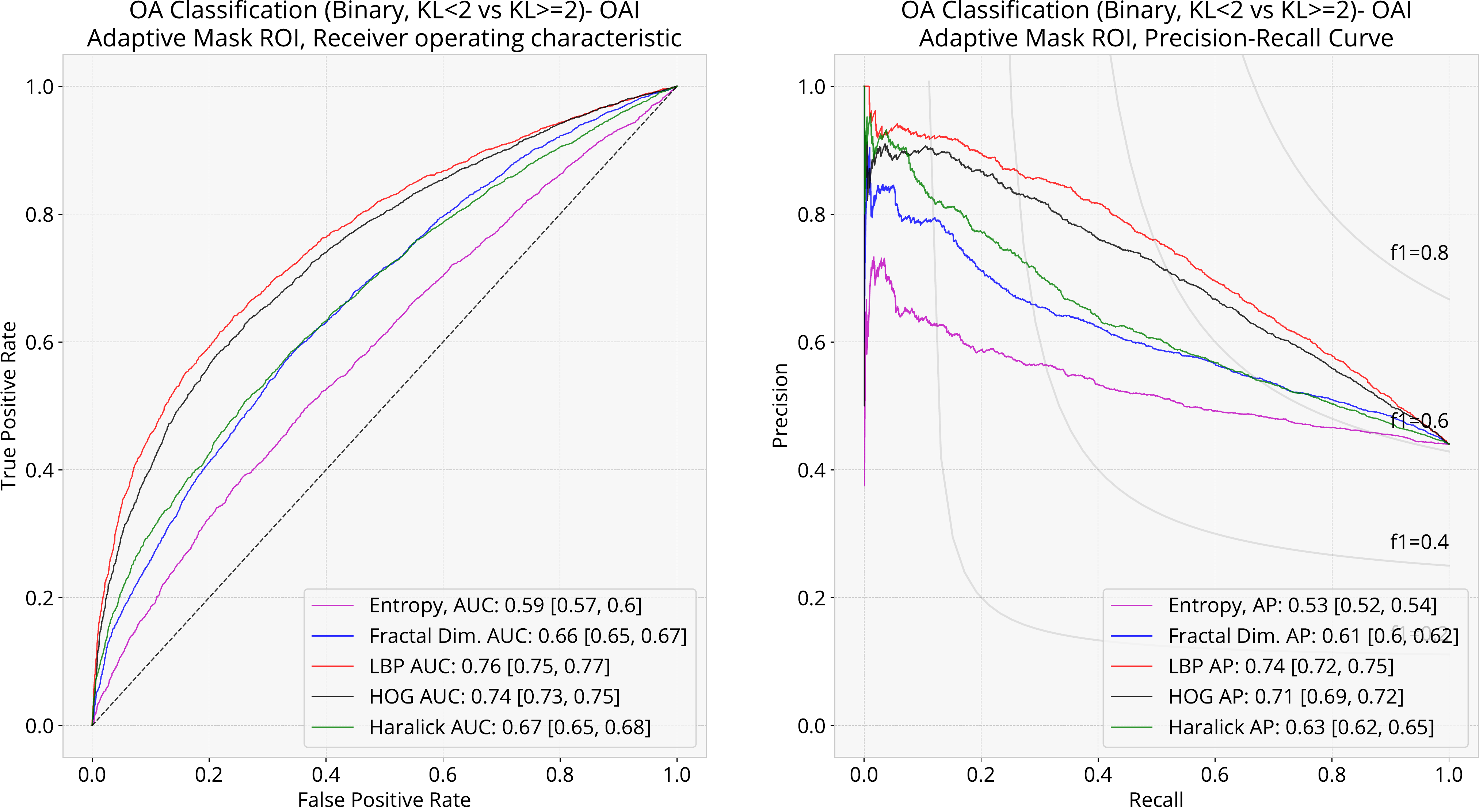}}
\caption{Classification performance of texture descriptors with {(a)} standard ROI and {(b)} adaptive mask ROI on OAI with logistic regression on 5-fold cross validation setting. In ROC plots, labels show AUC values and labels in PR curves show AP values with 95\% confidence intervals in parentheses.}
\label{fig: oai_curves_texture}
\end{figure}

\begin{figure}[!ht]
\centering 
\subfloat[]{\includegraphics[width = 1\linewidth]{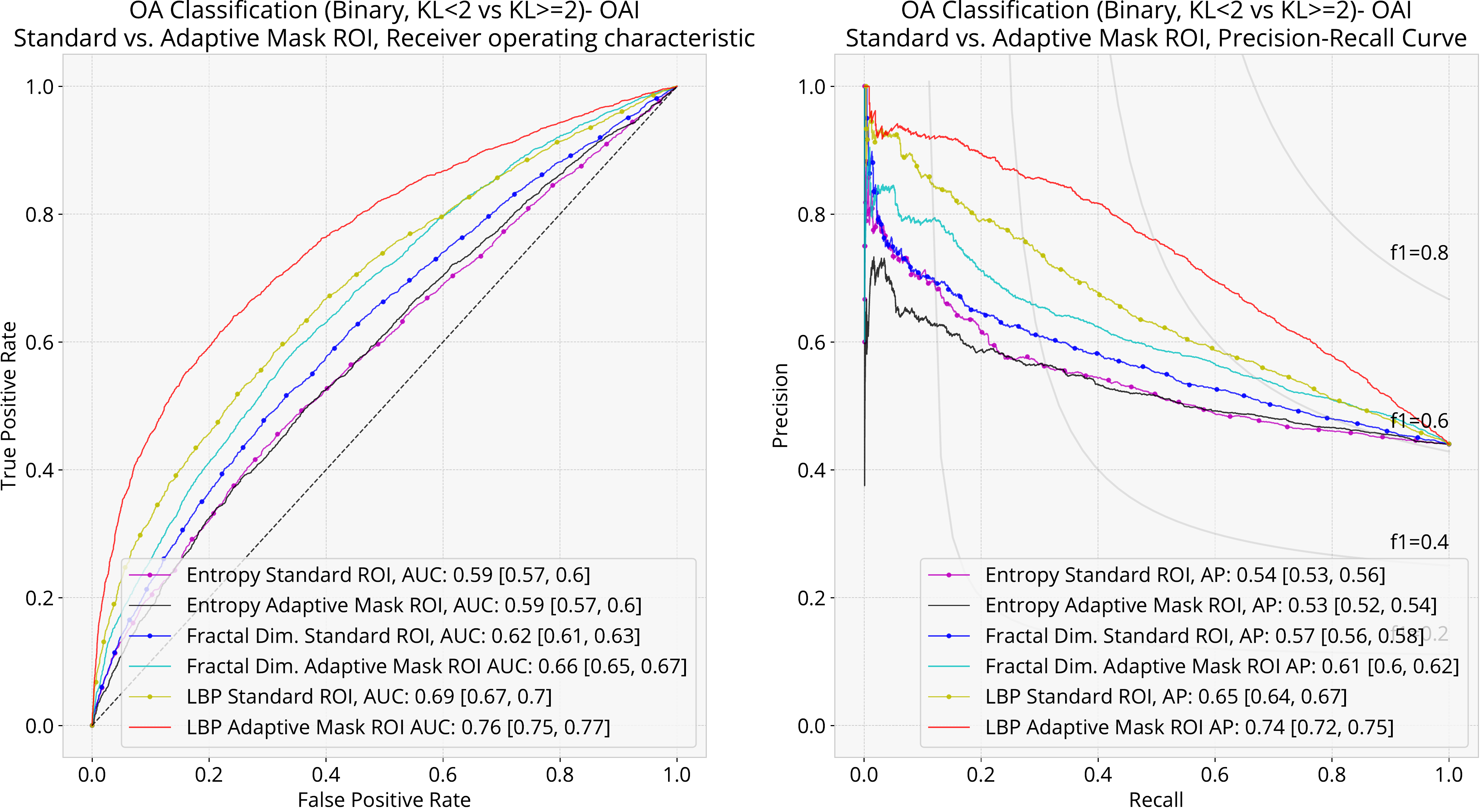}}
\hfill
\subfloat[]{\includegraphics[width = 1\linewidth]{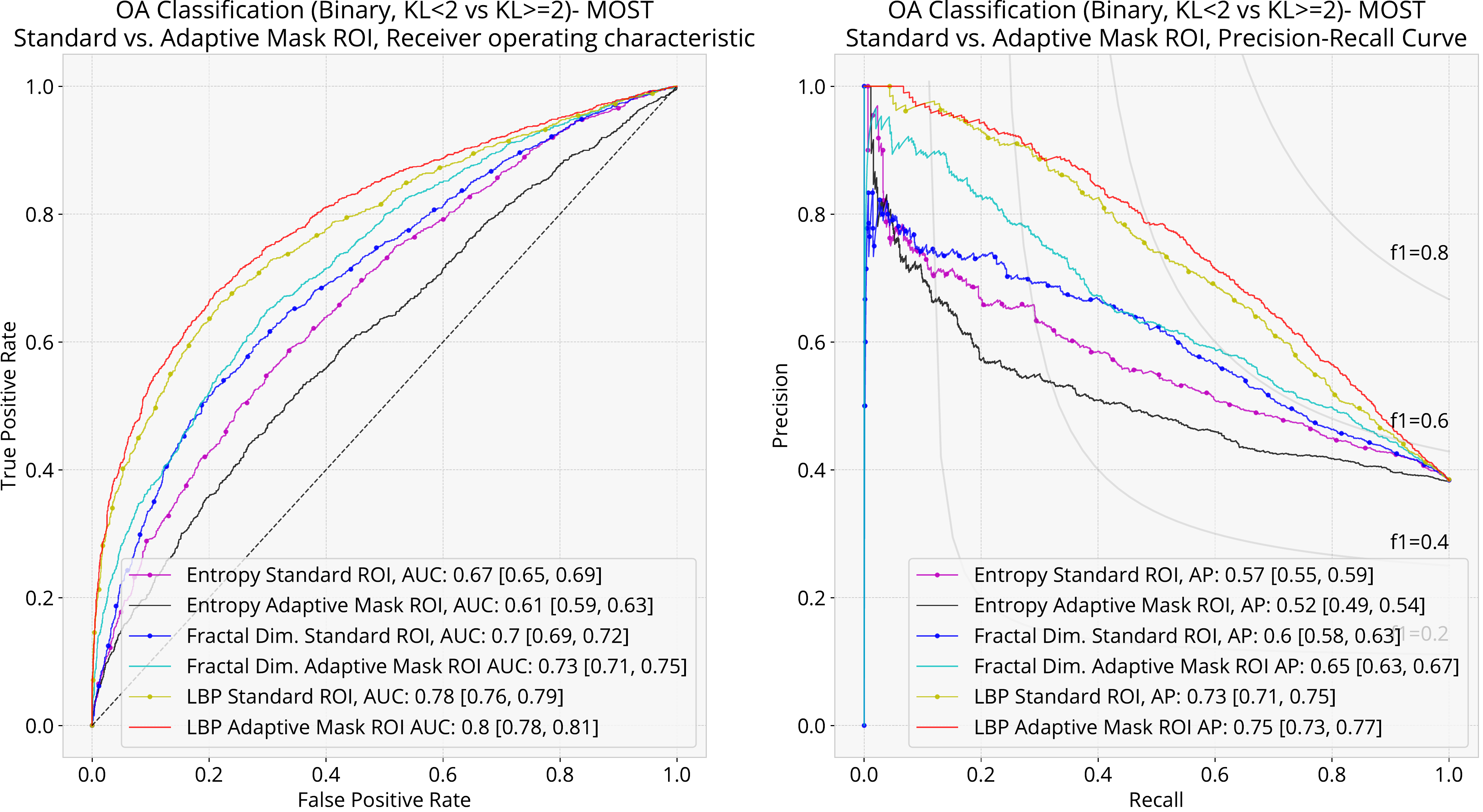}}
\caption{Comparison of standard ROI vs adaptive mask ROI with {(a)} OAI and {(b)} MOST using logistic regression on 5-fold cross validation setting. In ROC plots, labels show AUC values and labels in PR curves show AP values with 95\% confidence intervals in parentheses.}
\label{fig: compare_std_adaptive_curve}
\end{figure}

In Table \ref{tab:OAI_region_comparison}, we present the ROC AUC and PR AP values for OAI dataset where we used 5-fold cross validation.
From this table, it can be seen that the performance of the classifier increases when the texture descriptors are calculated from adaptive ROI instead of standard ROI.
The highest differences were observed with LBP (AUC $7.6\%$) and HOG (AUC $9.1\%$) descriptors.
We made a similar analysis for MOST dataset and presented the results in Table \ref{tab:most_region_comparison}.
Consistent with previous finding, both ROC AUC and PR AP scores for adaptive ROI are better except Entropy descriptor.
Figure \ref{fig: oai_curves_texture} shows the ROC and PR curves for OAI data.
The ROC and PR curves for MOST data are presented in \nameref{sec:supp} Figure \ref{fig: most_curves_texture}.
In addition, we demonstrated the effect of region selection (standard ROI vs. adaptive mask ROI) with ROC and PR curves. 
Figure \ref{fig: compare_std_adaptive_curve} demonstrates the ROC and PR curves for LBP, fractal dimension, and entropy descriptors for OAI dataset.

\subsubsection*{Independent Test Set}
In Table \ref{tab:independent_region_comparison}, we presented the results where we used the OAI data for training and the MOST data for testing (validation experiment). 
We also see from this table that utilizing adaptive mask (\hyperref[fig: region26]{\textcolor{magenta}{\texttt{t26}}}) clearly improves the classification performance over standard ROI for LBP, HOG, and Haralick descriptors where they provide higher classification rate compared to Fractal dimension and Entropy.
We also show the ROC and PR curves for the validation experiments in \nameref{sec:supp} Figure \ref{fig: validation_curves_texture}.

\subsubsection*{Feature and Region Combination}
We analyzed the effect of feature combination on tibial medial margin and also analyzed the region combination (combining features from medial margin and lateral margin) on the classification performance.   
\nameref{sec:supp} Table \ref{tab:combine_feature} shows ROC AUC and AP values for feature combination on region \hyperref[fig: region26]{\textcolor{magenta}{\texttt{t26}}}.
Combination of LBP with HOG, FD, and Haralick yielded AUC of 0.774 [0.764, 0.783] and AP of 0.754 [0.742, 0.765] on OAI which is only slightly higher than LBP alone. 
Similarly, for MOST, we found that feature combination of LBP and HOG which yielded the best performance with  AUC of 0.808 [0.793, 0.822] and  AP of 0.766 [0.747, 0.782] did not provide significant performance difference in AUC and in AP when compared to LBP alone.
In \nameref{sec:supp} Table \ref{tab:combine_region}, we presented classification performances for combining features from multiple ROIs from lateral and medial side of tibia (e.g. $LBP_{t26}+ LBP_{t3}$). 
Combination of LBP features from lateral and medial tibia margins yielded AUC of 0.791 [0.782, 0.801] and AP of 0.778 [0.767, 0.788] on OAI.
We observed similar performance increase with other descriptors (around AUC of 2\%) when combined from lateral and medial tibia regions for OAI and MOST as well.

%% file: sections/supplement.tex
\newcommand{\beginsupplement}{%
        \setcounter{table}{0}
        \renewcommand{\thetable}{S\arabic{table}}%
        \setcounter{figure}{0}
        \renewcommand{\thefigure}{S\arabic{figure}}%
     }
 
 \beginsupplement
 
 \section*{Supplementary}\label{sec:supp}

\subsection*{Parameter Optimization}
\begin{table}[!h]
\caption{Hyperparameter search space and best parameters found after performing grid search on 5-fold cross-validation setting.}
\label{tab:hyperparameter}
\centering
\begin{tabular}{lllp{4cm}l}
\toprule
                     & \textbf{Hyperparameter}    & \textbf{Type}        & \textbf{Values} & \textbf{Best} \\
                     \midrule
\multirow{2}{*}{LBP} & Radius            & Ordinal     & \{2,3,4,5,6\}   & 6       \\
                     & Points            & Ordinal     & \{8,10,12,16,24\}   & 8       \\
                     \midrule\\
\multirow{3}{*}{HOG} & Orientations      &     Ordinal        &   \{4, 8, 9 ,10\}     &   4      \\
                     & Cells per block &   Ordinal          &   \{(2,2), (3, 3), (4,4), (6,6)\}     &    \(4,4\)     \\
                     & Pixels per cell &             &       \{(6,6), (8,8), (10,10), (12,12), (14,14)\}    & \(10,10\) \\
                     \midrule
\end{tabular}
\end{table}

\newpage
\begin{figure}[!ht]
\null\hfill
\subfloat[]{\includegraphics[width = 0.8\linewidth]{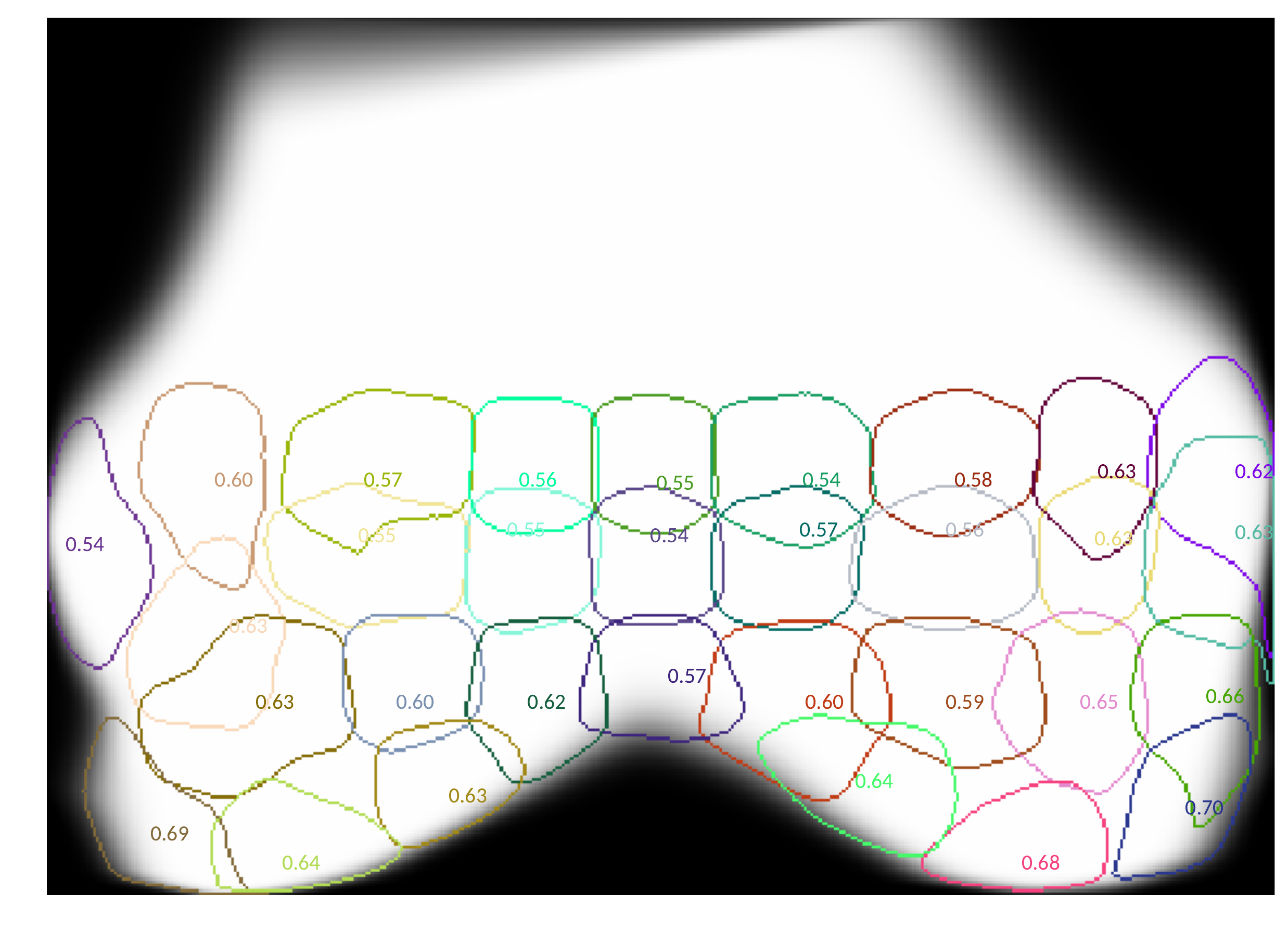}}
\hfill\null

\null\hfill
\subfloat[]{\includegraphics[width = 0.8\linewidth]{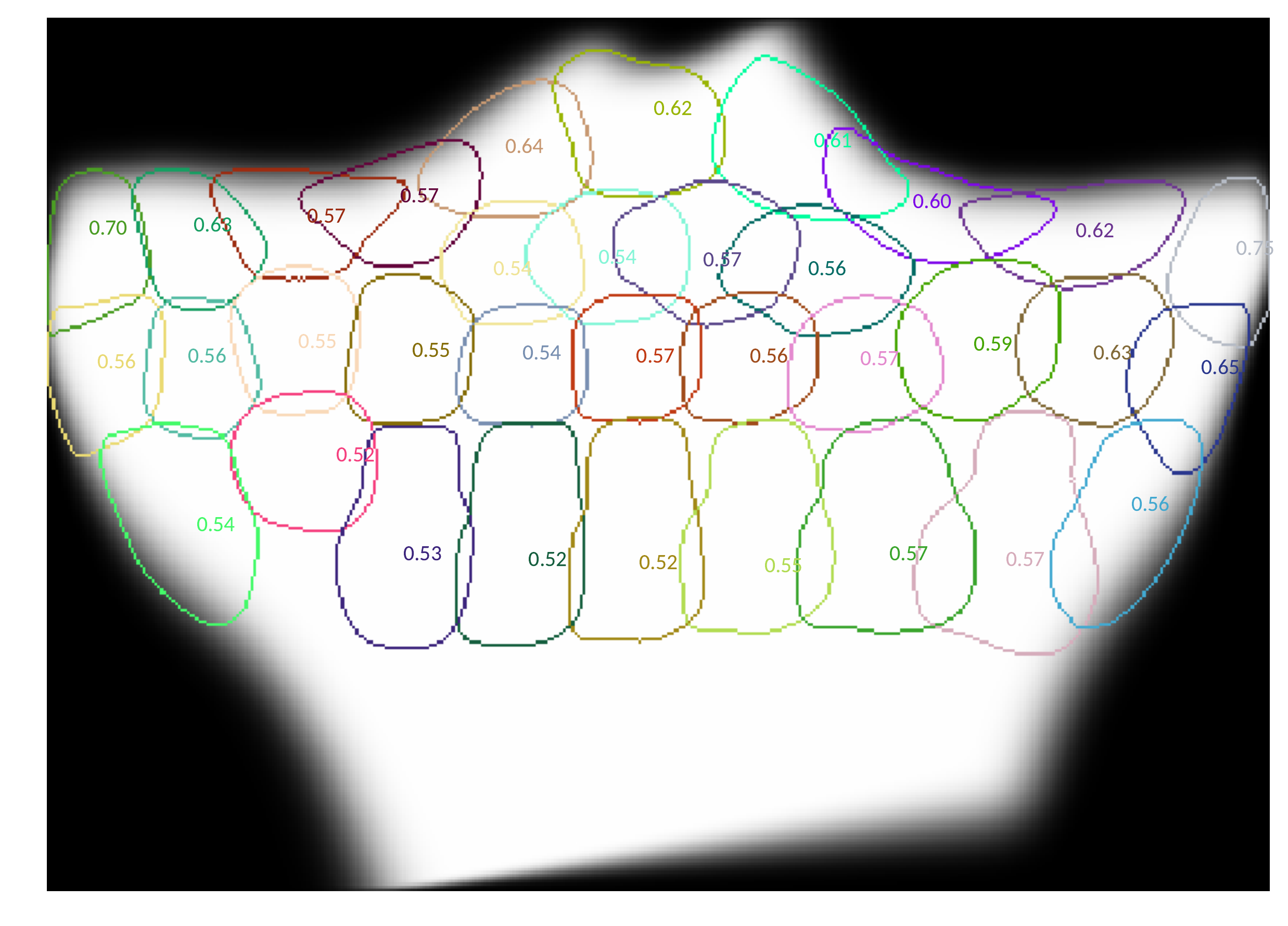}}\hfill\null

\caption{ This figure demonstrates several adaptive regions found by our framework on (a) femur and on (b) tibia. Contours represent the thresholded mask borders. ROC AUC values based on LBP features are also shown on the figure. Some of the (overlapping) regions are excluded for illustration reasons. Best viewed on screen.}
\label{fig: regions_summary}
\end{figure}

%%%%%%%%%%%%%%%%%%%%%%%%%%%%%%%%%%%%%%%%%%%%55

\begin{figure}[!ht]
\centering
\subfloat[]{\includegraphics[width = 1\linewidth]{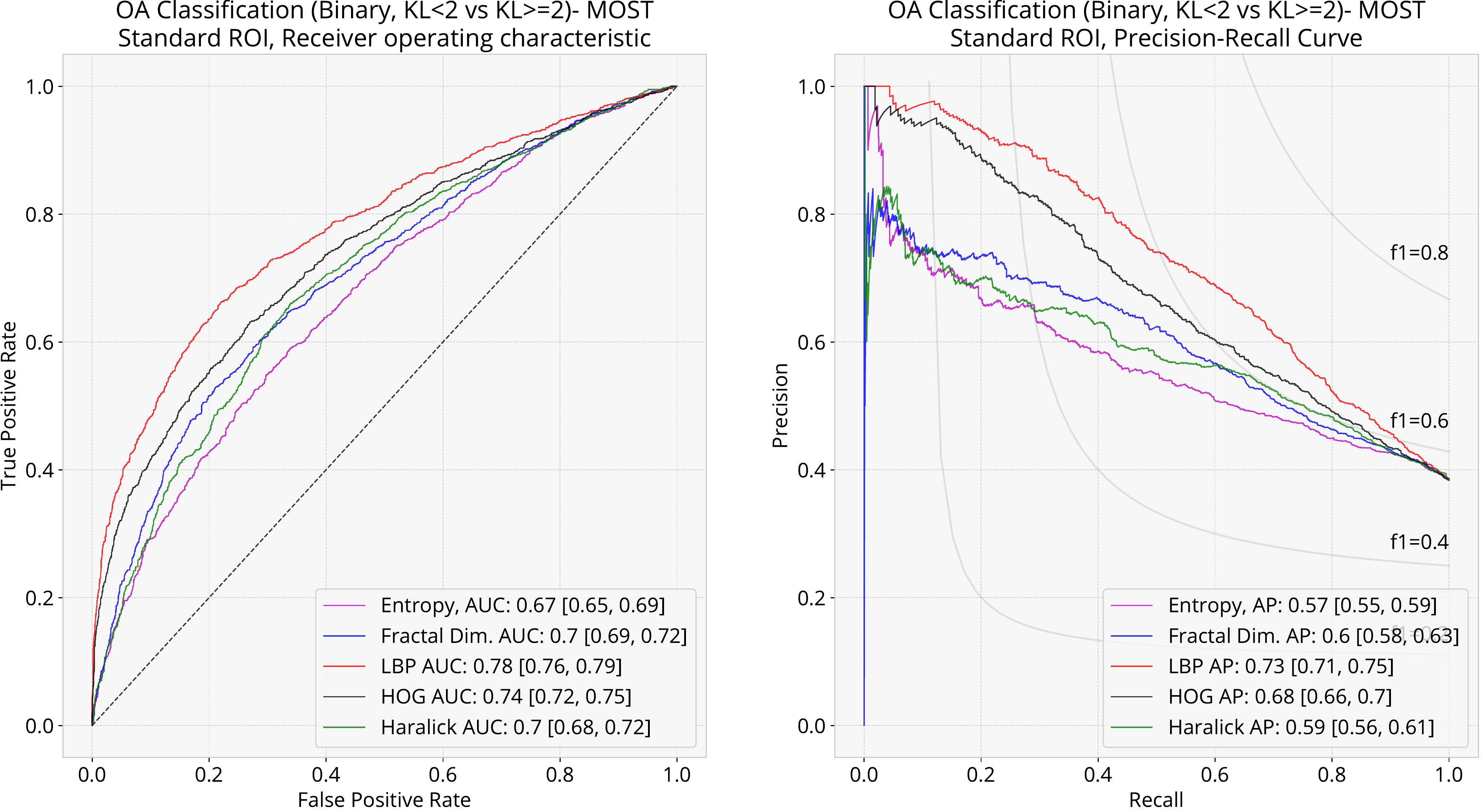}}
\hfill
\subfloat[]{\includegraphics[width = 1\linewidth]{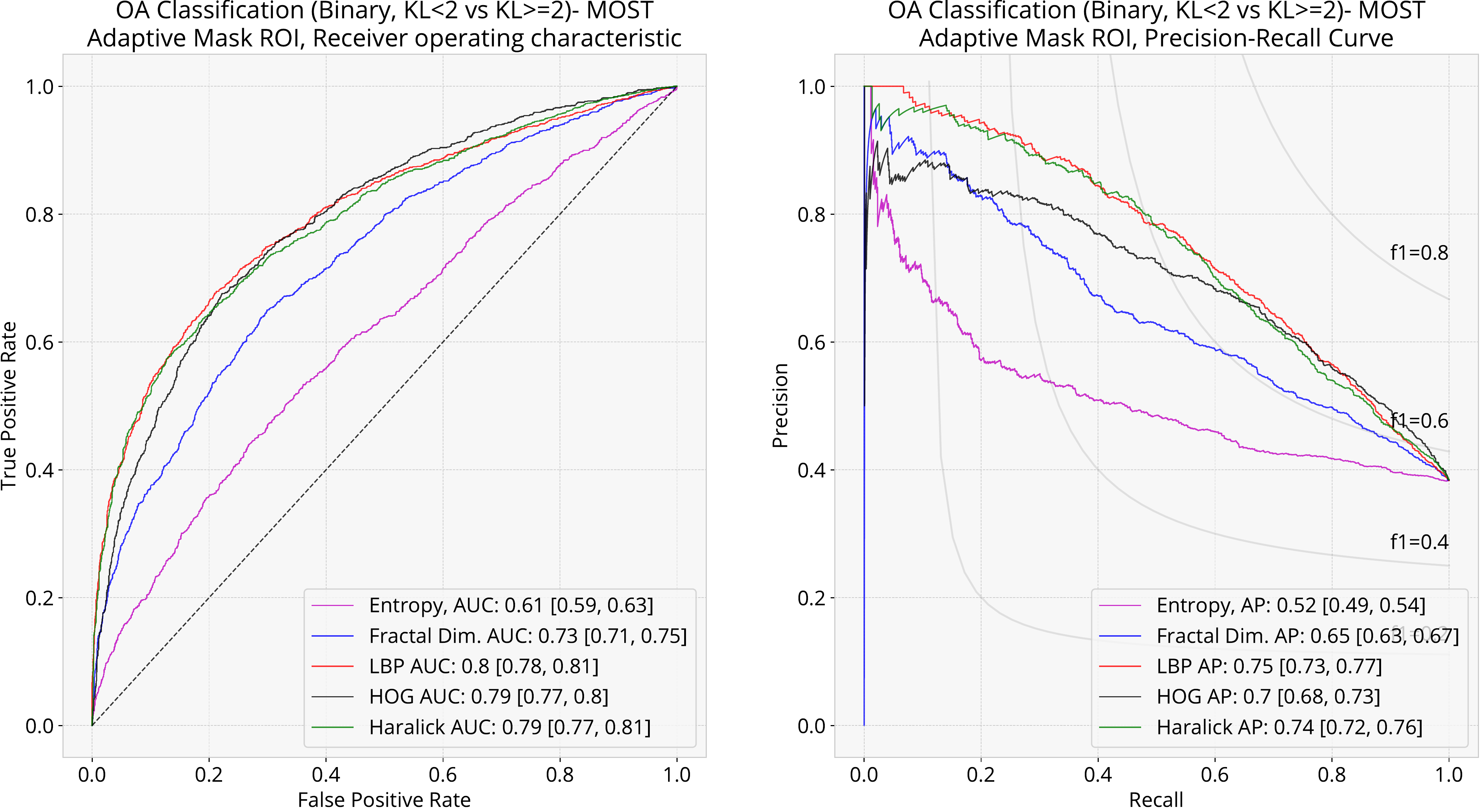}}
\caption{Classification performance of texture descriptors on standard ROI {(a)} and adaptive mask ROI {(b)} on \textbf{MOST} with Logistic regression on 5-fold cross validation setting. In ROC plots, labels show AUC values and the labels in PR curves show AP values with 95\% confidence intervals in parentheses. Best viewed on screen.}
\label{fig: most_curves_texture}
\end{figure}

\begin{figure}[!ht]
\centering
\subfloat[]{\includegraphics[width = 1\linewidth]{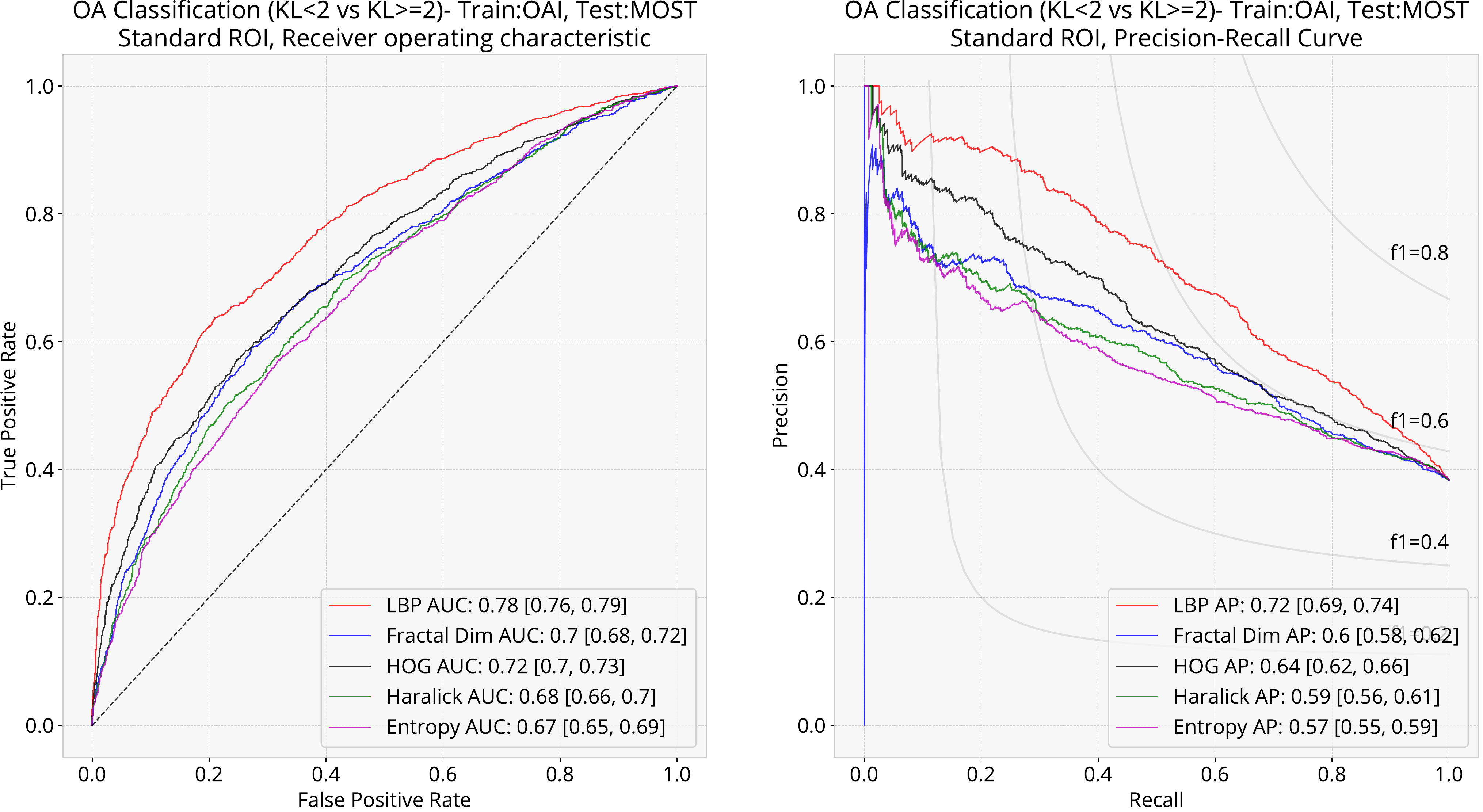}}
\hfill
\subfloat[]{\includegraphics[width = 1\linewidth]{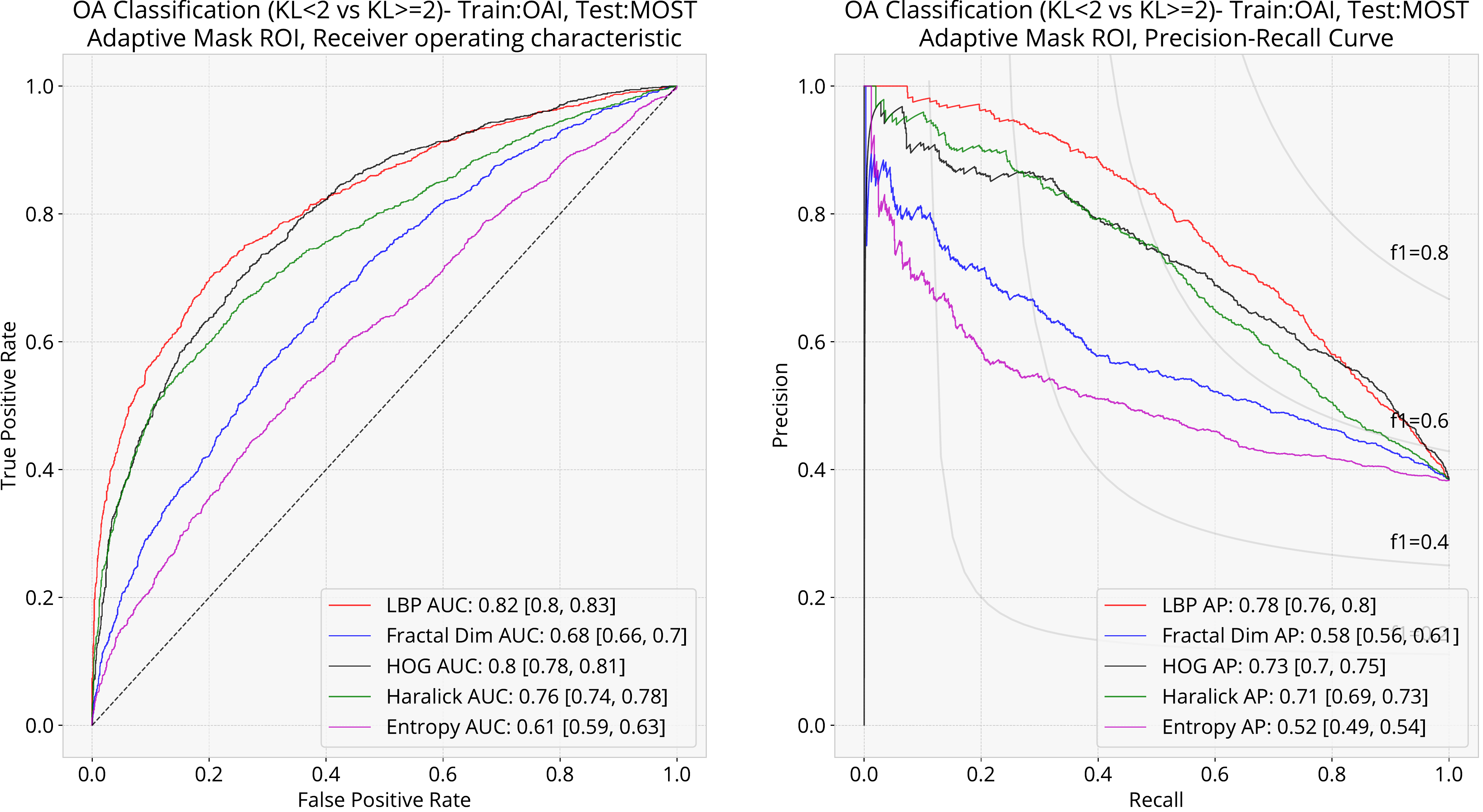}}
\caption{Classification performance of texture descriptors on standard ROI {(a)} and adaptive mask ROI {(b)} where we used \textbf{OAI as the training set} and \textbf{MOST data as a test set} (independent validation data). We used Logistic regression classification as before. In ROC plots, labels show AUC values and the labels in PR curves show AP values with 95\% confidence intervals in parentheses. Best viewed on screen.}
\label{fig: validation_curves_texture}
\end{figure}

%%%%%%%%%%%%%%%%%%%%%%%%%%%%%%%%%%%%%%%%%%%%%%%%%%%%%%5

\subfile{sections/supp_tables.tex}

%% file: sections/supp_tables.tex
%%%%%%%%%%%%%%%%%%%%%%%%%%%%%%%%%%%%%%%%%%%%%%%

\begin{sidewaystable}
\captionsetup{width=0.8\linewidth}
\caption{\small{Comparison of feature combinations computed at adaptive region \hyperref[fig: region26]{\textcolor{magenta}{\texttt{t26}}}. Scores are given by ROC AUC and Precision-Recall AP with 95\% confidence interval in parentheses.}}
    \label{tab:combine_feature}
\vspace{1cm}
\centering
\resizebox{\dimexpr 0.8\linewidth\relax}{!}{%
\begin{tabular}{ll|l|l|l}
\toprule
Method      & {Score}     &  \multicolumn{1}{c}{ \textbf{OAI - Exp 1}} & \multicolumn{1}{c}{ \textbf{MOST - Exp 2}} & \multicolumn{1}{p{3cm}}{\centering \textbf{Train: OAI,} \\ \textbf{Test: MOST - Exp 3}}  \\
\midrule
   LBP + HOG & AUC & 0.767 [0.757, 0.776] & 0.793 [0.777, 0.808]  & \textBf{0.822 [0.808, 0.835]}\\
    & AP &  0.745 [0.733, 0.756] & 0.738 [0.717, 0.756] &  0.766 [0.746, 0.784]\\
    LBP + HOG + Haralick & AUC & \textBf{0.774 [0.764, 0.783]} & \textBf{0.808 [0.793, 0.822]} & 0.813 [0.798, 0.827]\\
    & AP & \textBf{0.753 [0.741, 0.764]} & \textBf{0.766 [0.747, 0.782]} & \textBf{0.767 [0.747, 0.785]}\\
    LBP + HOG + Fractal  & AUC & 0.771 [0.760, 0.780] & 0.792 [0.776, 0.806]  & 0.805 [0.790, 0.819]\\
    & AP & 0.750 [0.738, 0.761] & 0.739 [0.717, 0.757] & 0.744 [0.723, 0.765] \\
    LBP + HOG + Fractal + Haralick & AUC &  \textBf{0.774 [0.764, 0.783]}  & 0.805 [0.790, 0.819] & 0.802 [0.786, 0.816]\\
    & AP & \textBf{0.754 [0.742, 0.765]} & {0.766 [0.747, 0.782]} & 0.754 [0.733, 0.773]\\
    % LBP + HOG + Fractal + \normalsize Haralick + Compact& AUC &  \textBf{0.786 [0.776, 0.795]}  & 0.805 [0.790, 0.819] & 0.811 [0.796, 0.825]\\
    % & AP & \textBf{0.772 [0.761, 0.782]} & 0.765 [0.746, 0.782] & 0.760 [0.740, 0.778]\\
    \bottomrule
    \end{tabular}
}
\end{sidewaystable}

%%%%%%%%%%%%%%%%%%%%%%%%%%%%%%%%%%%%%%%%%%%%%%%%%%%

\begin{sidewaystable}
\captionsetup{width=0.6\linewidth}
\caption{\small{Combining features from adaptive region \hyperref[fig: region26]{\textcolor{magenta}{\texttt{t26}}} (medial) and  \hyperref[fig: region]{\textcolor{cyan}{\texttt{t3}}} (lateral). Scores are given by ROC AUC and Precision-Recall AP with 95\% confidence interval in parentheses.}}
    \label{tab:combine_region}
\vspace{1cm}
\centerline{
\resizebox{\dimexpr 0.6\linewidth\relax}{!}{%
\begin{tabular}{ll|l|l|l}
\toprule
Method      & {Score}     &  \multicolumn{1}{c}{ \textbf{OAI - Exp 1}} & \multicolumn{1}{c}{ \textbf{MOST - Exp 2}} & \multicolumn{1}{p{3cm}}{\centering \textbf{Train: OAI,} \\ \textbf{Test: MOST - Exp 3}}  \\
\midrule
   $LBP_{t26}+ LBP_{t3}$  & AUC & 0.791 [0.782, 0.801] & 0.809 [0.793, 0.823]  & 0.840 [0.825, 0.852]\\
    & AP &  0.778 [0.767, 0.788] & 0.769 [0.750, 0.787] &  0.804 [0.786, 0.820]\\
    $HOG_{t26}+ HOG_{t3}$& AUC &  0.760 [0.749, 0.770] & 0.783 [0.767, 0.798] & 0.805 [0.789, 0.818]\\
    & AP & 0.734 [0.721, 0.745] & 0.691 [0.668, 0.714] & 0.740 [0.719, 0.758]\\
    $FD_{t26}+ FD_{t3}$  & AUC & 0.688 [0.677, 0.698] & 0.742 [0.725, 0.758] & 0.692 [0.674, 0.709]\\
    & AP & 0.646 [0.633, 0.658] & 0.669 [0.646, 0.690] &  0.588 [0.563, 0.610]\\
    $Haralick_{t26}+ Haralick_{t3}$ & AUC & 0.695 [0.684, 0.705] & 0.802 [0.785, 0.816]  & 0.759 [0.741, 0.775]\\
    & AP & 0.667 [0.654, 0.680] & 0.767 [0.748, 0.784] & 0.708 [0.685, 0.729] \\
    % Compact & AUC &  \textBf{0.838 [0.830, 0.846]}  &\textBf{0.904 [0.893, 0.913]} &\textBf{0.857 [0.843, 0.869]}\\
    % & AP &  \textBf{0.832 [0.823, 0.840]} & \textBf{0.874 [0.860, 0.886]} & \textBf{0.828 [0.811, 0.842]}\\
    \bottomrule
    \end{tabular}
}
}
\end{sidewaystable}

%% file: bibliography.bib
@inproceedings{ljuhar2018combining,
  title={Combining fractal-and entropy-based bone texture analysis for the prediction of Osteoarthritis: data from the Multicenter Osteoarthritis study (MOST)},
  author={Ljuhar, R and Bertalan, Z and Ljuhar, D and Fahrleitner-Pammer, A and Nehrer, S and Dimai, H},
  booktitle={R{\"o}Fo-Fortschritte auf dem Gebiet der R{\"o}ntgenstrahlen und der bildgebenden Verfahren},
  volume={190},
  number={S 01},
  pages={WISS--106},
  year={2018},
  organization={Georg Thieme Verlag KG}
}

@inproceedings{jennane2014variational,
  title={A variational model for trabecular bone radiograph characterization},
  author={Jennane, Rachid and Touvier, J{\'e}r{\^o}me and Bergounioux, Maitine and Lespessailles, Eric},
  booktitle={Biomedical Imaging (ISBI), 2014 IEEE 11th International Symposium on},
  pages={1283--1286},
  year={2014},
  organization={IEEE}
}

@article{messent2005tibial,
  title={Tibial cancellous bone changes in patients with knee osteoarthritis. A short-term longitudinal study using Fractal Signature Analysis},
  author={Messent, Elizabeth A and Ward, Rupert J and Tonkin, Carol J and Buckland-Wright, Christopher},
  journal={Osteoarthritis and cartilage},
  volume={13},
  number={6},
  pages={463--470},
  year={2005},
  publisher={Elsevier}
}

@article{hafezi2018new,
  title={New imaging modalities to predict and evaluate osteoarthritis progression},
  author={Hafezi-Nejad, Nima and Guermazi, Ali and Demehri, Shadpour and Roemer, Frank W},
  journal={Best Practice \& Research Clinical Rheumatology},
  year={2018},
  publisher={Elsevier}
}

@article{wolski2011trabecular,
  title={Trabecular bone texture detected by plain radiography and variance orientation transform method is different between knees with and without cartilage defects},
  author={Wolski, Marcin and Stachowiak, Gwidon W and Dempsey, Alasdair R and Mills, Peter M and Cicuttini, Flavia M and Wang, Yuanyuan and Stoffel, Karl K and Lloyd, David G and Podsiadlo, Pawel},
  journal={Journal of Orthopaedic Research},
  volume={29},
  number={8},
  pages={1161--1167},
  year={2011},
  publisher={Wiley Online Library}
}

@article{kraus2009trabecular,
  title={Trabecular morphometry by fractal signature analysis is a novel marker of osteoarthritis progression},
  author={Kraus, Virginia Byers and Feng, Sheng and Wang, ShengChu and White, Scott and Ainslie, Maureen and Brett, Alan and Holmes, Anthony and Charles, H Cecil},
  journal={Arthritis \& Rheumatism: Official Journal of the American College of Rheumatology},
  volume={60},
  number={12},
  pages={3711--3722},
  year={2009},
  publisher={Wiley Online Library}
}

@article{riad2018texture,
  title={Texture analysis using complex wavelet decomposition for knee osteoarthritis detection: Data from the osteoarthritis initiative},
  author={Riad, Rabia and Jennane, Rachid and Brahim, Abdelbasset and Janvier, Thomas and Toumi, Hechmi and Lespessailles, Eric},
  journal={Computers \& Electrical Engineering},
  volume={68},
  pages={181--191},
  year={2018},
  publisher={Elsevier}
}

@article{lynch1991analysis,
  title={Analysis of texture in macroradiographs of osteoarthritic knees, using the fractal signature},
  author={Lynch, JA and Hawkes, DJ and Buckland-Wright, JC},
  journal={Physics in Medicine \& Biology},
  volume={36},
  number={6},
  pages={709},
  year={1991},
  publisher={IOP Publishing}
}

@article{messent2006differences,
  title={Differences in trabecular structure between knees with and without osteoarthritis quantified by macro and standard radiography, respectively},
  author={Messent, EA and Ward, RJ and Tonkin, CJ and Buckland-Wright, C},
  journal={Osteoarthritis and cartilage},
  volume={14},
  number={12},
  pages={1302--1305},
  year={2006},
  publisher={Elsevier}
}

@inproceedings{antony2016quantifying,
  title={Quantifying radiographic knee osteoarthritis severity using deep convolutional neural networks},
  author={Antony, Joseph and McGuinness, Kevin and O'Connor, Noel E and Moran, Kieran},
  booktitle={Pattern Recognition (ICPR), 2016 23rd International Conference on},
  pages={1195--1200},
  year={2016},
  organization={IEEE}
}

@article{lynch1991robust,
  title={A robust and accurate method for calculating the fractal signature of texture in macroradiographs of osteoarthritic knees},
  author={Lynch, JA and Hawkes, DJ and Buckland-Wright, JC},
  journal={Medical Informatics},
  volume={16},
  number={2},
  pages={241--251},
  year={1991},
  publisher={Taylor \& Francis}
}

@inproceedings{janvier2015roi,
  title={ROI impact on the characterization of knee osteoarthritis using fractal analysis},
  author={Janvier, Thomas and Toumi, Hechmi and Harrar, Khaled and Lespessailles, Eric and Jennane, Rachid},
  booktitle={Image Processing Theory, Tools and Applications (IPTA), 2015 International Conference on},
  pages={304--308},
  year={2015},
  organization={IEEE}
}

@article{kraus2018predictive,
  title={Predictive validity of radiographic trabecular bone texture in knee osteoarthritis: the Osteoarthritis Research Society International/Foundation for the National Institutes of Health Osteoarthritis Biomarkers Consortium},
  author={Kraus, Virginia Byers and Collins, Jamie E and Charles, H Cecil and Pieper, Carl F and Whitley, Lawrence and Losina, Elena and Nevitt, Michael and Hoffmann, Steve and Roemer, Frank and Guermazi, Ali and others},
  journal={Arthritis \& Rheumatology},
  volume={70},
  number={1},
  pages={80--87},
  year={2018},
  publisher={Wiley Online Library}
}

@inproceedings{minciullo2016fully,
  title={Fully automated shape analysis for detection of osteoarthritis from lateral knee radiographs},
  author={Minciullo, Luca and Cootes, Tim},
  booktitle={Pattern Recognition (ICPR), 2016 23rd International Conference on},
  pages={3787--3791},
  year={2016},
  organization={IEEE}
}

@article{hirvasniemi2017differences,
  title={Differences in tibial subchondral bone structure evaluated using plain radiographs between knees with and without cartilage damage or bone marrow lesions-the Oulu Knee Osteoarthritis study},
  author={Hirvasniemi, Jukka and Thevenot, J{\'e}r{\^o}me and Guermazi, Ali and Podlipsk{\'a}, Jana and Roemer, Frank W and Nieminen, Miika T and Saarakkala, Simo},
  journal={European radiology},
  volume={27},
  number={11},
  pages={4874--4882},
  year={2017},
  publisher={Springer}
}

@article{hladuuvka2017femoral,
  title={Femoral ROIs and Entropy for Texture-based Detection of Osteoarthritis from High-Resolution Knee Radiographs},
  author={Hlad{\r{u}}vka, Ji\v r{\' i} and Phuong, Bui Thi Mai and Ljuhar, Richard and Ljuhar, Davul and Rodrigues, Ana M and Branco, Jaime C and Canh{\~a}o, Helena},
  journal={arXiv preprint arXiv:1703.09296},
  year={2017}
}

@article{shamir2009knee,
  title={Knee X-ray image analysis method for automated detection of Osteoarthritis},
  author={Shamir, Lior and Ling, Shari M and Scott Jr, William W and Bos, Angelo and Orlov, Nikita and Macura, Tomasz J and Eckley, D Mark and Ferrucci, Luigi and Goldberg, Ilya G},
  journal={IEEE Transactions on Biomedical Engineering},
  volume={56},
  number={2},
  pages={407--415},
  year={2009},
  publisher={IEEE}
}

@article{shamir2009early,
  title={Early detection of radiographic knee osteoarthritis using computer-aided analysis},
  author={Shamir, Lior and Ling, Shari M and Scott, William and Hochberg, Marc and Ferrucci, Luigi and Goldberg, Ilya G},
  journal={Osteoarthritis and Cartilage},
  volume={17},
  number={10},
  pages={1307--1312},
  year={2009},
  publisher={Elsevier}
}

@article{wolski2010differences,
  title={Differences in trabecular bone texture between knees with and without radiographic osteoarthritis detected by directional fractal signature method},
  author={Wolski, Marcin and Podsiadlo, Pawel and Stachowiak, GW and Lohmander, LS and Englund, Martin},
  journal={Osteoarthritis and cartilage},
  volume={18},
  number={5},
  pages={684--690},
  year={2010},
  publisher={Elsevier}
}

@article{woloszynski2012prediction,
  title={Prediction of progression of radiographic knee osteoarthritis using tibial trabecular bone texture},
  author={Woloszynski, Tomasz and Podsiadlo, Pawel and Stachowiak, GW and Kurzynski, M and Lohmander, LS and Englund, Martin},
  journal={Arthritis \& Rheumatism},
  volume={64},
  number={3},
  pages={688--695},
  year={2012},
  publisher={Wiley Online Library}
}

@inproceedings{thomson2015automated,
  title={Automated shape and texture analysis for detection of osteoarthritis from radiographs of the knee},
  author={Thomson, Jessie and O’Neill, Terence and Felson, David and Cootes, Tim},
  booktitle={International Conference on Medical Image Computing and Computer-Assisted Intervention},
  pages={127--134},
  year={2015},
  organization={Springer}
}

@article{hirvasniemi2014quantification,
  title={Quantification of differences in bone texture from plain radiographs in knees with and without osteoarthritis},
  author={Hirvasniemi, Jukka and Thevenot, J{\'e}r{\^o}me and Immonen, Ville and Liikavainio, Tuomas and Pulkkinen, Pasi and J{\"a}ms{\"a}, Timo and Arokoski, Jari and Saarakkala, Simo},
  journal={Osteoarthritis and cartilage},
  volume={22},
  number={10},
  pages={1724--1731},
  year={2014},
  publisher={Elsevier}
}

@article{podsiadlo2008automated,
  title={Automated selection of trabecular bone regions in knee radiographs},
  author={Podsiadlo, P and Wolski, M and Stachowiak, GW},
  journal={Medical physics},
  volume={35},
  number={5},
  pages={1870--1883},
  year={2008},
  publisher={Wiley Online Library}
}

@article{woloszynski2010signature,
  title={A signature dissimilarity measure for trabecular bone texture in knee radiographs},
  author={Woloszynski, Tomasz and Podsiadlo, Pawel and Stachowiak, GW and Kurzynski, M},
  journal={Medical physics},
  volume={37},
  number={5},
  pages={2030--2042},
  year={2010},
  publisher={Wiley Online Library}
}

@article{wolski2009directional,
  title={Directional fractal signature analysis of trabecular bone: evaluation of different methods to detect early osteoarthritis in knee radiographs},
  author={Wolski, Marcin and Podsiadlo, Pawel and Stachowiak, GW},
  journal={Proceedings of the Institution of Mechanical Engineers, Part H: Journal of Engineering in Medicine},
  volume={223},
  number={2},
  pages={211--236},
  year={2009},
  publisher={SAGE Publications Sage UK: London, England}
}

@article{podsiadlo2014trabecular,
  title={Trabecular bone texture detected by plain radiography is associated with an increased risk of knee replacement in patients with osteoarthritis: a 6 year prospective follow up study},
  author={Podsiadlo, P and Cicuttini, Flavia M and Wolski, M and Stachowiak, GW and Wluka, AE},
  journal={Osteoarthritis and cartilage},
  volume={22},
  number={1},
  pages={71--75},
  year={2014},
  publisher={Elsevier}
}

@article{anifah2013osteoarthritis,
  title={Osteoarthritis classification using self organizing map based on gabor kernel and contrast-limited adaptive histogram equalization},
  author={Anifah, Lilik and Purnama, I Ketut Eddy and Hariadi, Mochamad and Purnomo, Mauridhi Hery},
  journal={The open biomedical engineering journal},
  volume={7},
  pages={18},
  year={2013},
  publisher={Bentham Science Publishers}
}

@article{podsiadlo2016baseline,
  title={Baseline trabecular bone and its relation to incident radiographic knee osteoarthritis and increase in joint space narrowing score: directional fractal signature analysis in the MOST study},
  author={Podsiadlo, P and Nevitt, MC and Wolski, Marcin and Stachowiak, GW and Lynch, JA and Tolstykh, I and Felson, DT and Segal, NA and Lewis, CE and Englund, M},
  journal={Osteoarthritis and cartilage},
  volume={24},
  number={10},
  pages={1736--1744},
  year={2016},
  publisher={Elsevier}
}

@article{janvier2017subchondral,
  title={Subchondral tibial bone texture predicts the incidence of radiographic knee osteoarthritis: data from the Osteoarthritis Initiative},
  author={Janvier, T and Jennane, R and Toumi, H and Lespessailles, E},
  journal={Osteoarthritis and cartilage},
  volume={25},
  number={12},
  pages={2047--2054},
  year={2017},
  publisher={Elsevier}
}

@article{brahim2019decision,
  title={A Decision Support Tool For Early Detection of Knee OsteoArthritis using X-ray Imaging and Machine Learning: Data from the OsteoArthritis Initiative},
  author={Brahim, Abdelbasset and Jennane, Rachid and Riad, Rabia and Janvier, Thomas and Khedher, Laila and Toumi, Hechmi and Lespessailles, Eric},
  journal={Computerized Medical Imaging and Graphics},
  year={2019},
  publisher={Elsevier}
}

@article{achanta2012slic,
  title={SLIC superpixels compared to state-of-the-art superpixel methods},
  author={Achanta, Radhakrishna and Shaji, Appu and Smith, Kevin and Lucchi, Aurelien and Fua, Pascal and S{\"u}sstrunk, Sabine and others},
  journal={IEEE transactions on pattern analysis and machine intelligence},
  volume={34},
  number={11},
  pages={2274--2282},
  year={2012},
  publisher={Institute of Electrical and Electronics Engineers, Inc., 345 E. 47 th St. NY~…}
}

@article{buckland2004subchondral,
  title={Subchondral bone changes in hand and knee osteoarthritis detected by radiography},
  author={Buckland-Wright, Christopher},
  journal={Osteoarthritis and cartilage},
  volume={12},
  pages={10--19},
  year={2004},
  publisher={Elsevier}
}

@article{kamibayashi1995trabecular,
  title={Trabecular microstructure in the medial condyle of the proximal tibia of patients with knee osteoarthritis},
  author={Kamibayashi, L and Wyss, UP and Cooke, TDV and Zee, B},
  journal={Bone},
  volume={17},
  number={1},
  pages={27--35},
  year={1995},
  publisher={Elsevier}
}

@article{lowitz2014characterization,
  title={Characterization of knee osteoarthritis-related changes in trabecular bone using texture parameters at various levels of spatial resolution—a simulation study},
  author={Lowitz, Torsten and Museyko, Oleg and Bousson, Valerie and Kalender, Willi A and Laredo, Jean Denis and Engelke, Klaus},
  journal={BoneKEy reports},
  volume={3},
  year={2014},
  publisher={Nature Publishing Group}
}

@inproceedings{neogi2012clinical,
  title={Clinical significance of bone changes in osteoarthritis},
  author={Neogi, Tuhina},
  booktitle={Arthritis Research \& Therapy},
  volume={14},
  number={2},
  pages={A3},
  year={2012},
  organization={BioMed Central}
}

@article{lindner2013fully,
  title={Fully automatic segmentation of the proximal femur using random forest regression voting},
  author={Lindner, Claudia and Thiagarajah, Shankhar and Wilkinson, J Mark and Wallis, Gillian A and Cootes, Timothy F and arcOGEN Consortium and others},
  journal={IEEE transactions on medical imaging},
  volume={32},
  number={8},
  pages={1462--1472},
  year={2013},
  publisher={IEEE}
}

@article{hirvasniemi2016correlation,
  title={Correlation of subchondral bone density and structure from plain radiographs with micro computed tomography ex vivo},
  author={Hirvasniemi, Jukka and Thevenot, J{\'e}r{\^o}me and Kokkonen, Harri T and Finnil{\"a}, Mikko A and Ven{\"a}l{\"a}inen, Mikko S and J{\"a}ms{\"a}, Timo and Korhonen, Rami K and T{\"o}yr{\"a}s, Juha and Saarakkala, Simo},
  journal={Annals of biomedical engineering},
  volume={44},
  number={5},
  pages={1698--1709},
  year={2016},
  publisher={Springer}
}

@inproceedings{dalal2005histograms,
  title={Histograms of oriented gradients for human detection},
  author={Dalal, Navneet and Triggs, Bill},
  booktitle={international Conference on computer vision \& Pattern Recognition (CVPR'05)},
  volume={1},
  pages={886--893},
  year={2005},
  organization={IEEE Computer Society}
}

@inproceedings{ojala2000gray,
  title={Gray scale and rotation invariant texture classification with local binary patterns},
  author={Ojala, Timo and Pietik{\"a}inen, Matti and M{\"a}enp{\"a}{\"a}, Topi},
  booktitle={European Conference on Computer Vision},
  pages={404--420},
  year={2000},
  organization={Springer}
}

@article{coelho2012mahotas,
  title={Mahotas: Open source software for scriptable computer vision},
  author={Coelho, Luis Pedro},
  journal={arXiv preprint arXiv:1211.4907},
  year={2012}
}

@article{tiulpin2019multimodal,
  title={Multimodal Machine Learning-based Knee Osteoarthritis Progression Prediction from Plain Radiographs and Clinical Data},
  author={Tiulpin, Aleksei and Klein, Stefan and Bierma-Zeinstra, Sita and Thevenot, J{\'e}r{\^o}me and Rahtu, Esa and van Meurs, Joyce and Oei, Edwin HG and Saarakkala, Simo},
  journal={arXiv preprint arXiv:1904.06236},
  year={2019}
}

@article{tiulpin2018automatic,
  title={Automatic knee osteoarthritis diagnosis from plain radiographs: a deep learning-based approach},
  author={Tiulpin, Aleksei and Thevenot, J{\'e}r{\^o}me and Rahtu, Esa and Lehenkari, Petri and Saarakkala, Simo},
  journal={Scientific reports},
  volume={8},
  number={1},
  pages={1727},
  year={2018},
  publisher={Nature Publishing Group}
}

@article{hirvasniemi2019bone,
  title={Bone Density and Texture from Minimally Post-Processed Knee Radiographs in Subjects with Knee Osteoarthritis},
  author={Hirvasniemi, Jukka and Niinim{\"a}ki, Jaakko and Thevenot, J{\'e}r{\^o}me and Saarakkala, Simo},
  journal={Annals of biomedical engineering},
  pages={1--10},
  year={2019},
  publisher={Springer}
}

@article{egloff2012biomechanics,
  title={Biomechanics and pathomechanisms of osteoarthritis},
  author={Egloff, Christian and H{\"u}gle, Thomas and Valderrabano, Victor},
  journal={Swiss medical weekly},
  volume={142},
  number={2930},
  year={2012},
  publisher={EMH Media}
}

@article{veenland1996estimation,
  title={Estimation of fractal dimension in radiographs},
  author={Veenland, JF and Grashuis, JL and van der Meer, Frits and Beckers, ALD and Gelsema, ES},
  journal={Medical physics},
  volume={23},
  number={4},
  pages={585--594},
  year={1996},
  publisher={Wiley Online Library}
}

@article{kadir2001saliency,
  title={Saliency, scale and image description},
  author={Kadir, Timor and Brady, Michael},
  journal={International Journal of Computer Vision},
  volume={45},
  number={2},
  pages={83--105},
  year={2001},
  publisher={Springer}
}

@article{norman2019applying,
  title={Applying densely connected convolutional neural networks for staging osteoarthritis severity from plain radiographs},
  author={Norman, Berk and Pedoia, Valentina and Noworolski, Adam and Link, Thomas M and Majumdar, Sharmila},
  journal={Journal of digital imaging},
  volume={32},
  number={3},
  pages={471--477},
  year={2019},
  publisher={Springer}
}

@misc{tiulpin2019kneel,
    title={KNEEL: Knee Anatomical Landmark Localization Using Hourglass Networks},
    author={Aleksei Tiulpin and Iaroslav Melekhov and Simo Saarakkala},
    year={2019},
    eprint={1907.12237},
    archivePrefix={arXiv},
    primaryClass={cs.CV}
}

@inproceedings{varma2007locally,
  title={Locally invariant fractal features for statistical texture classification},
  author={Varma, Manik and Garg, Rahul},
  booktitle={2007 IEEE 11th international conference on computer vision},
  pages={1--8},
  year={2007},
  organization={IEEE}
}

@article{scikit-learn,
 title={Scikit-learn: Machine Learning in {P}ython},
 author={Pedregosa, F. and Varoquaux, G. and Gramfort, A. and Michel, V.
         and Thirion, B. and Grisel, O. and Blondel, M. and Prettenhofer, P.
         and Weiss, R. and Dubourg, V. and Vanderplas, J. and Passos, A. and
         Cournapeau, D. and Brucher, M. and Perrot, M. and Duchesnay, E.},
 journal={Journal of Machine Learning Research},
 volume={12},
 pages={2825--2830},
 year={2011}
}
